\documentclass[default]{ancillary/sn-jnl}

\usepackage{amsmath,amssymb,amstext}
\usepackage{bookmark}
\usepackage[T1]{fontenc}
\usepackage{graphicx}
\definecolor{ForestGreen}{HTML}{2e8b21}
\definecolor{NatureRed}{HTML}{f54e42}
\definecolor{NatureBlue}{HTML}{4254f5}
\definecolor{NatureGreen}{HTML}{138f57}
\definecolor{NatureMagenta}{HTML}{6b1345}
\definecolor{NaturePurple}{HTML}{c227b0}
\definecolor{NatureCyan}{HTML}{2b807b}
\definecolor{NatureOrange}{HTML}{f5a142}
\usepackage{hyperref}
\hypersetup{
    colorlinks=true,
    linkcolor=blue,
    citecolor=blue,
    filecolor=magenta,      
    urlcolor=cyan,
    }
\usepackage{subcaption}
\usepackage{txfonts}
\usepackage{textgreek}
\usepackage{xargs}
\usepackage{xcolor}
\usepackage{xspace}

\graphicspath{{./}{figs/}}

\jyear{2023}

\usepackage{etoolbox}
\makeatletter
\newcommand\sendemail[3]{
\edef\@tempa{mailto:#1?subject=#2 }%
\edef\@tempb{\expandafter\html@spaces\@tempa\@empty}%
\href{\@tempb}{#3}}

\catcode\%=11
\def\html@spaces#1 #2{#1
\catcode\%=14
\makeatother

\newcommand{\citationneeded}{\textcolor{ForestGreen}{$^{\rm citation\;needed}$}}
\let\oldtextsigma\textsigma
\renewcommand{\textsigma}{\oldtextsigma\xspace}
\let\oldAA\AA
\renewcommand{\AA}{\text{\oldAA}\xspace}
\let\oldtextdegree\textdegree
\renewcommand{\textdegree}{\oldtextdegree\xspace}
\def\w80{\ensuremath{w_{80}}\xspace}

\newcommand{\kms}{\ensuremath{\mathrm{km\,s^{-1}}}\xspace}
\newcommand{\MSun}{\ensuremath{{\rm M}_\odot}\xspace}
\newcommand{\yr}{\ensuremath{{\rm yr}}\xspace}
\newcommand{\Myr}{\ensuremath{{\rm Myr}}\xspace}
\newcommand{\Gyr}{\ensuremath{{\rm Gyr}}\xspace}
\newcommand{\peryr}{\ensuremath{{\rm yr^{-1}}}\xspace}
\newcommand{\Lsun}{\hbox{\,${\rm L}_\odot$}}
\newcommand{\mum}{\text{\textmu m}\xspace}
\newcommand{\kpc}{\text{kpc}\xspace}
\newcommand{\ZH}{\text{[Z/H]}\xspace}

\newcommandx{\lambdar}[2][1=R,2=]{\ensuremath{\lambda_{\rm {#1}}{#2}}\xspace}
\newcommand{\eps}{\ensuremath{\epsilon}\xspace}
\newcommand{\Mstar}{\ensuremath{M_\star}\xspace}
\newcommand{\Mdyn}{\ensuremath{M_\mathrm{dyn}}\xspace}
\newcommand{\re}{\ensuremath{R_\mathrm{e}}\xspace}
\newcommand{\vstar}{\ensuremath{v_\star}\xspace}
\newcommand{\vnai}{\ensuremath{v_{\NaI}}\xspace}
\newcommand{\sigmastar}{\ensuremath{\sigma_\star}\xspace}
\newcommand{\sigmaestar}{\ensuremath{\sigma_{\star,\mathrm{e}}}\xspace}
\newcommand{\vperc}[1]{\ensuremath{v_{#1}}\xspace}

\newcommand{\vesc}{\ensuremath{v_\mathrm{esc}}\xspace}
\newcommand{\nelec}{\ensuremath{n_\mathrm{e}}\xspace}
\newcommand{\Rout}{\ensuremath{R_\mathrm{out}}\xspace}
\newcommand{\vout}{\ensuremath{v_\mathrm{out}}\xspace}
\newcommandx{\Mout}[2][1=,2=]{\ensuremath{M_{\mathrm{out}{#2}}^{#1}}\xspace}
\newcommandx{\Mdotout}[2][1=,2=]{\ensuremath{\dot{M}_{\mathrm{out}{#2}}^{#1}}\xspace}

\newcommandx{\fluxdcgs}[1][1=-20]{$\times 10^{[#1]}$~erg~s$^{-1}$~cm$^{-2}$~\AA$^{-1}$\xspace}
\newcommandx{\powercgs}[1][1=44]{$\times 10^{#1}$~erg~s$^{-1}$\xspace}
\newcommand{\Av}{\ensuremath{A_V}\xspace}

\newcommand{\hda}{\ensuremath{\mathrm{H\text{\textdelta}_A}}\xspace}
\newcommand{\hga}{\ensuremath{\mathrm{H\text{\textgamma}_A}}\xspace}
\newcommand{\Halpha}{\text{H\textalpha}\xspace}
\newcommand{\Hbeta}{\text{H\textbeta}\xspace}
\newcommand{\Hgamma}{\text{H\textgamma}\xspace}
\newcommand{\Hdelta}{\text{H\textdelta}\xspace}
\newcommand{\Pabeta}{\text{Pa\textbeta}\xspace}
\newcommand{\Hepsilon}{\text{H\textepsilon}\xspace}
\newcommandx{\permittedEL}[6][1=O,2=III,3=,4=,5=,6=]{\text{{#1}\,{\sc {#2}}{#3}{#4}{#5}{#6}}\xspace}
\newcommandx{\semiforbiddenEL}[6][1=O,2=III,3=,4=,5=,6=]{\text{{#1}\,{\sc{#2}}]{#3}{#4}{#5}{#6}}\xspace}
\newcommandx{\forbiddenEL}[6][1=O,2=III,3=,4=,5=,6=]{\text{[{#1}\,{\sc{#2}}]{#3}{#4}{#5}{#6}}\xspace}

\newcommand{\CaII}{\permittedEL[Ca][ii]}
\newcommand{\OIII}{\forbiddenEL[O][iii]}
\newcommandx{\OIIIL}[1][1=5007]{\forbiddenEL[O][iii][\textlambda][#1]}
\newcommand{\OIIIall}{\forbiddenEL[O][iii][\textlambda][\textlambda][4960,][5008]}
\newcommandx{\NIL}{\forbiddenEL[N][i][\textlambda][5201]}
\newcommand{\OI}{\forbiddenEL[O][i]}
\newcommand{\OIall}{\forbiddenEL[O][i][\textlambda][\textlambda][6302,][6366]}
\newcommand{\HeI}{\permittedEL[He][i]}
\newcommand{\NaI}{\permittedEL[Na][i]}
\newcommand{\NII}{\forbiddenEL[N][ii]}
\newcommandx{\NIIL}[1][1=6584]{\forbiddenEL[N][ii][\textlambda][#1]}
\newcommand{\NIIall}{\forbiddenEL[N][ii][\textlambda][\textlambda][6550,][6585]}
\newcommand{\SII}{\forbiddenEL[S][ii]}
\newcommand{\SIIall}{\forbiddenEL[S][ii][\textlambda][\textlambda][6718,][6733]}

\newcommandx{\target}[1][1=]{\text{GS-10578{#1}}\xspace}

\newcommand{\jwst}{\textit{JWST}\xspace}
\newcommand{\hst}{\textit{HST}\xspace}
\newcommand{\ppxf}{{\sc ppxf}\xspace}

\newcommand{\Mdynvalue}{$\Mdyn = 2.0\pm0.5 \times 10^{11}$~\MSun}

\raggedbottom

\begin{document}

\title[AGN feedback in PSB \target]{A fast-rotator post-starburst galaxy quenched by supermassive black-hole feedback at z=3}

\author*[1,2]{\fnm{Francesco} \sur{D'Eugenio}}\email{francesco.deugenio@gmail.com}\equalcont{These authors contributed equally to this work.}
\author[3]{\fnm{Pablo~G.} \sur{P\'erez-Gonz\'alez}}
\equalcont{These authors contributed equally to this work.}
\author[1,2,4]{\fnm{Roberto} \sur{Maiolino}}
\author[1,2]{\fnm{Jan} \sur{Scholtz}}
\author[3]{\fnm{Michele} \sur{Perna}}
\author[5,4]{\fnm{Chiara} \sur{Circosta}}
\author[1,2]{\fnm{Hannah} \sur{\"Ubler}}
\author[3]{\fnm{Santiago} \sur{Arribas}}
\author[6]{\fnm{Torsten} \sur{B\"{o}ker}}
\author[7]{\fnm{Andrew~J.} \sur{Bunker}}
\author[8]{\fnm{Stefano} \sur{Carniani}}
\author[9]{\fnm{Stephane} \sur{Charlot}}
\author[7]{\fnm{Jacopo} \sur{Chevallard}}
\author[10]{\fnm{Giovanni} \sur{Cresci}}
\author[11]{\fnm{Emma} \sur{Curtis-Lake}$^{\textit{\normalfont{11}}}$} 
\author[7]{\fnm{Gareth~C.} \sur{Jones}}
\author[12]{\fnm{Nimisha} \sur{Kumari}}
\author[3]{\fnm{Isabella} \sur{Lamperti}}
\author[1,2]{\fnm{Tobias~J.} \sur{Looser}}
\author[8]{\fnm{Eleonora} \sur{Parlanti}}
\author[13]{\fnm{Hans-Walter} \sur{Rix}}
\author[14]{\fnm{Brant} \sur{Robertson}}
\author[3]{\fnm{Bruno} \sur{Rodr\'iguez Del~Pino}}
\author[1,2]{\fnm{Sandro} \sur{Tacchella}}
\author[8]{\fnm{Giacomo} \sur{Venturi}}
\author[15]{\fnm{Chris~J.} \sur{Willott}}

\affil[1]{Kavli Institute for Cosmology, University of Cambridge, Madingley Road, Cambridge, CB3 OHA, UK}

\affil[2]{Cavendish Laboratory - Astrophysics Group, University of Cambridge, 19 JJ Thomson Avenue, Cambridge, CB3 OHE, UK}

\affil[3]{Centro de Astrobiolog\'{\i}a (CAB), CSIC-INTA, Ctra. de Ajalvir km 4, Torrej\'on de Ardoz, E-28850, Madrid, Spain}

\affil[4]{Department of Physics and Astronomy, University College London, Gower Street, London WC1E 6BT, UK}

\affil[5]{European Space Agency (ESA), European Space Astronomy Centre (ESAC), Camino Bajo del Castillo s/n, 28692 Villanueva de la Cañada, Madrid, Spain}

\affil[6]{European Space Agency, c/o STScI, 3700 San Martin Drive, Baltimore, MD 21218, USA}

\affil[7]{University of Oxford, Department of Physics, Denys Wilkinson Building, Keble Road, Oxford OX13RH, United Kingdom}

\affil[8]{Scuola Normale Superiore, Piazza dei Cavalieri 7, I-56126 Pisa, Italy}

\affil[9]{Sorbonne Universit\'e, UPMC-CNRS, UMR7095, Institut d'Astrophysique de Paris, F-75014 Paris, France}

\affil[10]{INAF - Osservatorio Astrofisico di Arcetri, largo E. Fermi 5, 50127 Firenze, Italy}

\affil[11]{Centre for Astrophysics Research, Department of Physics, Astronomy and Mathematics, University of Hertfordshire, Hatfield AL10 9AB, UK}

\affil[12]{AURA for European Space Agency, Space Telescope Science Institute, 3700 San Martin Drive. Baltimore, MD, 21210}

\affil[13]{Max Planck Institute for Astronomy, K\"onigstuhl 17, 69117 Heidelberg, Germany}

\affil[14]{Department of Astronomy and Astrophysics University of California, Santa Cruz, 1156 High Street, Santa Cruz CA 96054, USA}

\affil[15]{NRC Herzberg, 5071 West Saanich Rd, Victoria, BC V9E 2E7, Canada}

\abstract{
    There is compelling evidence that the most massive galaxies in 
    the Universe stopped forming stars due to the time-integrated feedback from their 
    central super-massive black holes (SMBHs). However, the exact quenching
    mechanism is not yet understood, because local massive galaxies were quenched
    billions of years ago.
    We present \jwst/NIRSpec integral-field spectroscopy
    observations of \target, a massive, quiescent galaxy at redshift $z=3.064$.
    From the spectrum we infer that the galaxy has a stellar mass of \Mstar = $1.6\pm0.2 \times 10^{11}$~\MSun
    and a dynamical mass \Mdynvalue. Half of its stellar mass formed at
    $z=3.7\text{--}4.6$, and the system is now quiescent,
    with the current star-formation rate SFR $<9$~\MSun~\peryr.
    We detect ionised- and neutral-gas outflows traced by \OIII emission and
    \NaI absorption. Outflow velocities reach $\vout \approx 1,000~\kms$,
    comparable to the galaxy escape velocity and too high to be explained
    by star formation alone. \target hosts an Active Galactic Nucleus (AGN),
    evidence that these outflows are due to SMBH feedback.
    The outflow rates are $0.14\text{--}2.9$ and $30\text{--}300~\MSun~\peryr$ for the ionised
    and neutral phases, respectively. The neutral outflow rate is ten times higher
    than the SFR, hence this is direct evidence for ejective SMBH feedback, with mass-loading
    capable of interrupting star formation by rapidly removing its fuel.
    Stellar kinematics show ordered rotation, with spin parameter
    $\lambdar[\re]=0.62\pm0.07$, meaning \target is rotation supported.
    This study shows direct evidence for ejective AGN feedback in a
    massive, recently quenched galaxy, thus clarifying how SMBHs quench their hosts.
    Quenching can occur without destroying the stellar disc.
}

\maketitle

\begin{figure}
  \centering
  \includegraphics[width=1\textwidth]{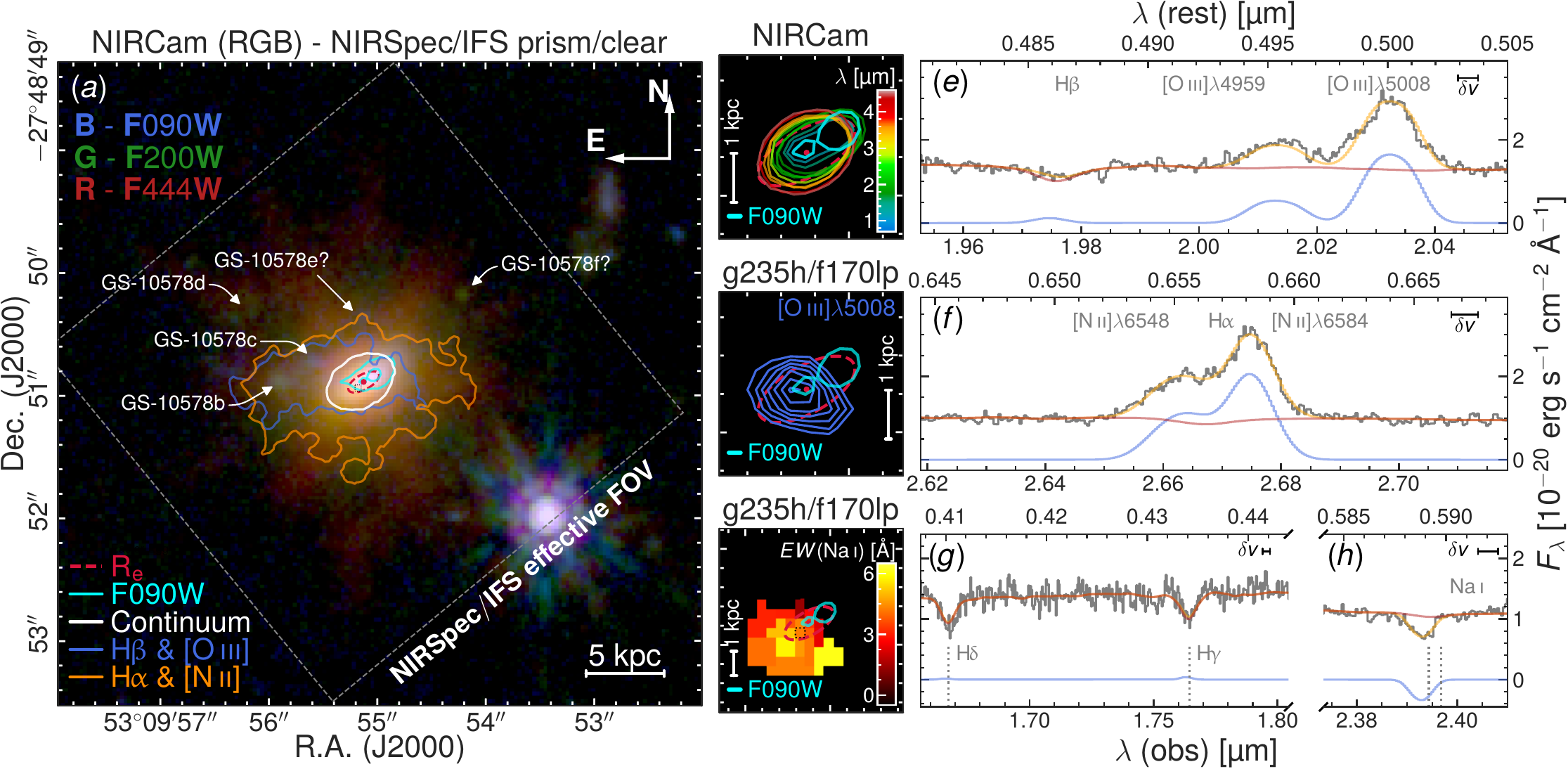}
  {\phantomsubcaption\label{f.emlines.a}
   \phantomsubcaption\label{f.emlines.b}
   \phantomsubcaption\label{f.emlines.c}
   \phantomsubcaption\label{f.emlines.d}
   \phantomsubcaption\label{f.emlines.e}
   \phantomsubcaption\label{f.emlines.f}
   \phantomsubcaption\label{f.emlines.g}
   \phantomsubcaption\label{f.emlines.h}
   }
  \caption{\textbf{Panel~\subref{f.emlines.a}.} False-colour image of \target from NIRCam. The grey dashed square
  is the field of view of NIRSpec/IFS (with dithering).
  The solid white contour is a stellar isophote from NIRSpec/IFS prism spectroscopy, the red
  dashed ellipse is the best-fit deconvolved model, with semi-major axis equal to 1~\re. The
  cyan contours are rest-frame UV from NIRCam F090W, displaying complex morphology,
  probably due to foreground dust.
  The orange/blue contours
  are isosignificance lines for the \Halpha--\NII and \Hbeta--\OIII blended lines,
  from the prism data. The nebular emission is significantly more extended than the stellar
  light, indicating the ability of mergers/feedback to enrich the circum-galactic medium (CGM).
  \textbf{Panel~\subref{f.emlines.b}.} Comparing the model isophote
  (red dashed ellipse) to NIRCam isophote morphology, there is a clear wavelength-dependent offset
  to the north west.
  \textbf{Panel~\subref{f.emlines.c}.} Brightest \OIII emission from g235h/f170lp data, showing a slight offset from the centre of the stellar emission (cf. red dot, distance $<1$~spaxel)
  and a remarkably different position angle, almost \textpi/2 compared the stars.
  \textbf{Panel~\subref{f.emlines.d}.} Neutral gas \NaI absorption, which is clearly asymmetric
  and located where the rest-UV photometry is faintest.
  \textbf{Panels~\subref{f.emlines.e}--\subref{f.emlines.h}.} Spectrum from the
  dotted black spaxel highlighted in panels~\subref{f.emlines.a} and~\subref{f.emlines.d},
  showing the data (grey), best-fit continuum and nebular-line model (red and blue, respectively) and the
  best-fit model (orange); the horizontal bar in the top right corner
  of each panel is $\delta v = 500 \, \kms$. Panels~\subref{f.emlines.e} and~\subref{f.emlines.f} show the
  complex, multi-component emission-line profile of \OIII and $\Halpha\text{--}\NII$.
  Panels \subref{f.emlines.g} and~\subref{f.emlines.h} show the same spaxel.
  The vertical dotted lines mark the
  position of the stellar \Hdelta, \Hgamma and \NaI-doublet absorption; the deep \NaI absorption, blueshifted with respect to the stars,
  betrays the presence of a neutral-gas outflow.
  }\label{f.emlines}
\end{figure}

The Universe today is not what it used to be \citep{madau+dickinson2014}. Local, 
massive quiescent galaxies stand like
colossal wrecks of glorious but remote star-formation histories, and mighty and rapid
quenching the likes of which have no present-day equals \citep{thomas+2010, mcdermid+2015}.
\jwst enables us for the first time to witness these monumental galaxies during the long-gone
epoch when they arose and fell.
By redshift $z=1.5\text{--}2$, 3--4~Gyr after the Big Bang, massive quiescent galaxies
have little to no cold gas, the fuel for star formation \citep{whitaker+2021, williams_alma_2021}.
But the question whether the missing fuel was consumed by starbursts or if it was removed
by `ejective' feedback from SMBHs remains open \citep{man+belli_quenching_2018, somerville+dave2015}.
We present a NIRSpec/IFS impression of \target, a massive, quiescent
galaxy at redshift $z=3.064$ (look-back time 11.7~Gyr). Observed as part of the
Galaxy Assembly with NIRSpec Integral Field Spectroscopy GTO programme (GA-NIFS),
this galaxy  (Fig.~\ref{f.emlines.a}) was selected as a `blue
nugget' \citep{barro+2013}, a class of massive, extremely compact 
galaxies (stellar mass $\Mstar=3\text{--}30\times10^{10}~\MSun$, half-light radius 
$\re=0.5\text{--}2~\kpc$), thought to
be the progenitors of compact quiescent galaxies at $z=2$ (`red nuggets', \citep{damjanov+2009}).
Blue nuggets are believed to be undergoing `gas-rich compaction', that is, a central 
starburst driven by disc instability or gas-rich major mergers, followed by rapid 
quenching \citep{dekel+2014} and leaving a compact, quiescent red-nugget galaxy.
As we will show, \target is instead already a red nugget in an advanced stage of 
quenching.
The system is merging with multiple low-mass satellites and is undergoing powerful, 
ejective feedback from its SMBH.

NIRCam imaging \citep{eisenstein_overview_2023, rieke_jades_2023, williams_jems_2023} reveals a
system with several smaller companions, all within 8~kpc in projection (labelled \target[b]--f
in~Fig.~\ref{f.emlines.a}, of which \target[b]--d are spectroscopically confirmed to be satellites). Multiple
low-mass satellites are a common occurrence for massive quiescent galaxies at high redshift 
\citep{suess_buddies_2023}.
\target shows a regular, elliptical shape at 2.77--4.44~\mum, but the light distribution is increasingly
lopsided blueward of 2~\mum
(Fig.~\ref{f.emlines.b}), with the brightest region at 0.9~\mum
clearly offset to the north west (by 0.5~\kpc) and displaying two peaks.

\begin{figure}
\centering
  \includegraphics[width=1\textwidth]{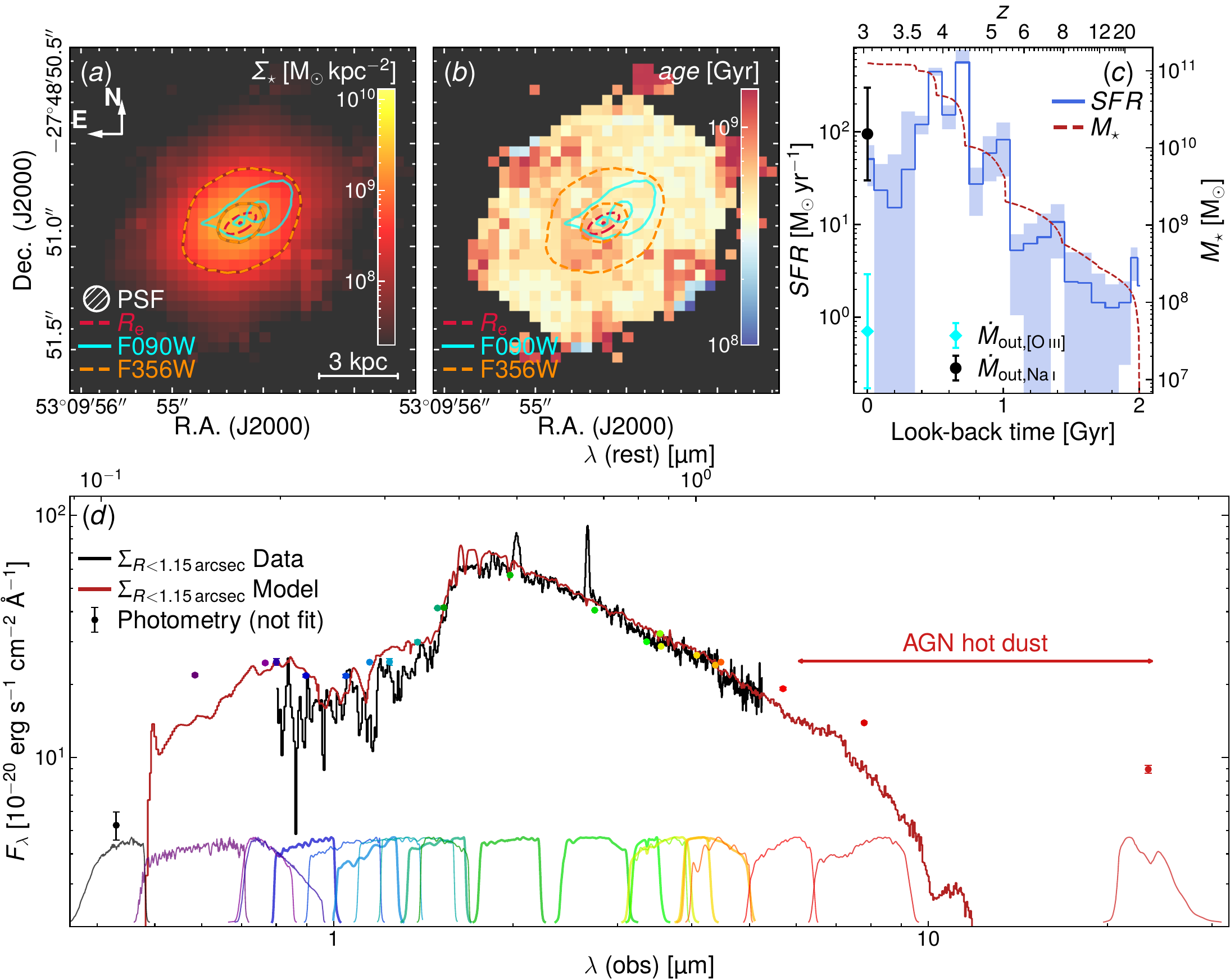}
  {\phantomsubcaption\label{f.sfh.a}
   \phantomsubcaption\label{f.sfh.b}
   \phantomsubcaption\label{f.sfh.c}
   \phantomsubcaption\label{f.sfh.d}}
  \caption{
  \textbf{Panels~\subref{f.sfh.a}--\subref{f.sfh.b}.} Stellar-population
  surface mass-density and mass-weighted age maps from full spectral fitting of
  the low-resolution spectroscopy data, showing that most of \target consists of
  0.5-Gyr-old stars. Merging satellites have lower mass-to-light ratios than the
  \target, and do not contribute to the mass-weighted properties.
  \textbf{Panel~\subref{f.sfh.c}.} The star-formation history from the same models, 
  integrated over the 2-d map, shows the main episode of star formation occurred 
  0.5~Gyr ago. The shaded region is the 1-\textsigma uncertainty. The cyan
  diamond and black circle are the mass outflow rates from \OIII and \NaI,
  respectively. Unlike the ionised-gas outflow, the neutral-gas outflow 
  has mass loading high enough to stop star formation.
  \textbf{Panel~\subref{f.sfh.d}.} SED of the individual spaxels summed inside a circular aperture of radius 1.15~arcsec, for both the NIRSpec/IFS data (black)
  and the best-fit model (red). We also show the aperture photometry as errorbars.
  The model does not include an AGN, therefore it does not capture the emission
  lines or the flux redder than 6~\mum, which is dominated by host-dust emission
  from the AGN torus.
  }
\label{f.sfh}
\end{figure}

We use the NIRSpec/IFS data with the low-resolution prism disperser to perform
spaxel-by-spaxel full spectral energy distribution (SED) modelling, to measure the
surface stellar mass density, spatially resolved stellar population properties and
star-formation history (SFH; see Methods). The resulting mass distribution is
symmetric, with no strong signs of asymmetry (Fig.~\ref{f.sfh.a}). The asymmetric 
light distribution at rest-frame UV wavelengths 
arises from younger stellar populations (luminous but with low stellar mass), and/or 
from an asymmetric dust distribution.
Integrating the 2-d properties inside a circular aperture of 1.2~arcsec radius,
and correcting for aperture losses, we measure a total stellar
mass $\Mstar = 1.6\pm0.2\times 10^{11}~\MSun$ (surviving stellar mass).
The total SFH from all spaxels shows that the main star-formation episode 
happened at $z=3.7\text{--}4.6$, 0.4--0.8~Gyr
prior to observation.
Afterwards, the star-formation rate (SFR) rapidly declined. There is some evidence of 
a recent upturn, but our models do not include an Active Galactic Nucleus (AGN), 
therefore this upturn could be due to the nebular emission being interpreted as due to 
star formation.
The SFR averaged over the last 100~Myr, is $40\pm20~\MSun~\peryr$.
This is five times below the star-forming main sequence at $z=3$
\citep{popesso+2023}, and comparable to the quiescence threshold of
19~\MSun~\peryr \citep{pacifici+2016}. However, part of the SFR is
likely due to the satellites, and the nebular emission is mostly powered by the 
AGN, meaning that the true SFR is likely much lower.
This implies that \target is currently quenched and on its way to quiescence,
which is consistent with the empirical UVJ-colour diagnostic \citep{williams+2009},
and with the SFR inferred from the emission-line analysis.

The NIRSpec/IFS observation in the low spectral-resolution configuration reveals 
extended 
nebular emission in both the \Hbeta--\OIII and \Halpha--\NII spectrally blended
complexes (Fig.~\ref{f.emlines.a}).
This nebular emission is elongated in a different direction compared to the stellar
major axis, but in the same direction as \target[b], which suggests that the ionised 
gas comes from inter-stellar medium (ISM) material stripped by the
interaction with the satellite. An outflow origin
is also possible, but in this case, the alignment with the satellite would be a
coincidence.

\begin{figure}
  \centering
  \includegraphics[width=.8\textwidth]{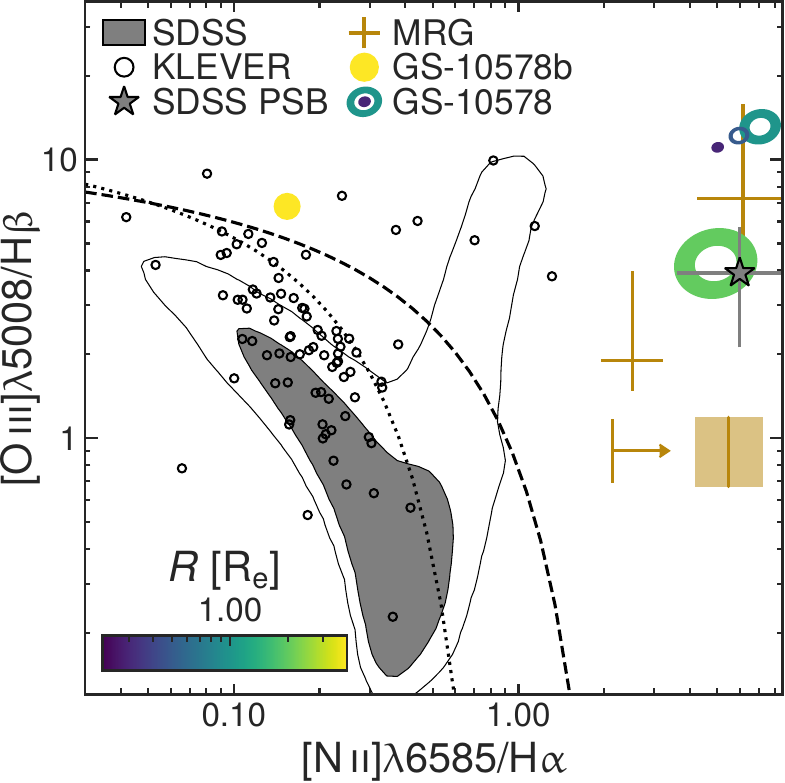}
  \caption{Emission-line diagnostic diagram for \target compared to 
  selected samples from the literature; the line ratios of our target place 
  its ionisation source firmly beyond the range that can be explained by star 
  formation (dashed line, \citep{kauffmann+2003c}).
  For the main target, we show the emission-line ratios from spectra integrated
  inside elliptical apertures (colour coded by the semi-major axis of the
  aperture). The yellow circle is the eastern satellite. We show a number
  of samples for comparison: the contours and the small empty circles trace the
  distribution of galaxies at $z=0.1$ (from SDSS) and $z=2$ (from KLEVER,
  \citep{curti+2020}), respectively. The
  sand symbols and upper limits are the MRG sample of post-starbust galaxies at $z=1\text{--}2$
  \citep{newman+2018a}, while the
  grey star is one of the most extreme post-starburst galaxies in SDSS. The
  extreme values we find for \target are similar to what is found for other
  post-starburst galaxies.}\label{f.bpt}
\end{figure}

Unlike the low-resolution observations, NIRSpec high-resolution data
(Fig.~\ref{f.emlines.b}--\subref{f.emlines.c} and~\subref{f.emlines.e}--\subref{f.emlines.f})
are able to separate \Halpha from \NII, and \Hbeta (which in \target we observe mostly in absorption) from \OIII. The emission lines show a
broad, multi-component profile, indicating complex kinematics. The BPT diagnostic
diagram (Fig.~\ref{f.bpt}) shows that \target has everywhere a high \NII/\Halpha ratio
(emission-line ratios are measured inside concentric elliptical annuli, following
the shape and position angle of the stellar isophotes). From the outmost to the innermost
annulus around \target, \OIIIL/\Hbeta increases from 4 to 10, whereas \NIIL/\Halpha remains
approximately constant at 5--8. All annuli are in the AGN region of the diagram, well beyond the 
demarcation curve between starburst-driven and AGN-driven photoionisation
\citep{kauffmann+2003c};
moreover, all annuli lie far outside the 99\textsuperscript{th} percentile of the 
distribution of local galaxies and AGN (grey and black contours).
Star-forming and AGN galaxies at $z=2$ also have
much smaller \NIIL/\Halpha ratios (empty circles), but we find our values to be similar to
post-starburst galaxies (PSB) at $z=1\text{--}2.5$ \citep{newman+2018b} and to the most extreme
PSBs in the local Universe (grey star).
Assuming all \Halpha emission was due to star formation, the
extinction-corrected SFR on timescales of 3--10~Myr is 9~\MSun~\peryr (inside 2~\re, where
\re=1.1~kpc from \citep{vanderwel+2014}). Varying the aperture size, the SFR
ranges between 4~\MSun~\peryr (inside 1~\re) to 19~\MSun~\peryr (summing over the entire map,
including the satellites). These values are similar to the result from the SED analysis.
However, as we have seen, \Halpha 
is dominated by emission from the AGN, therefore the true SFR is likely much lower, and
we treat all these values as upper limits within their apertures. Overall, the low \Halpha SFR
values are in agreement with the quiescent interpretation.
In addition to the BPT diagram, direct evidence for an AGN comes from the X-ray 
detection ($L_X = 8$~\powercgs[44] \citep{luo+2017}) and from the SED excess at
MIR wavelengths, which can be explained by the presence of a dusty torus around the accreting
SMBH (Fig.~\ref{f.sfh.d}, cf.~model and photometry at 24~\mum). Modelling the photometry using both
stellar and AGN emission models gives an AGN bolometric luminosity $L_\mathrm{AGN}=2.6\pm0.9$~\powercgs[45] (see Methods; \citep{circosta+2019}).

\begin{figure}
  \centering
  \includegraphics[width=1\textwidth]{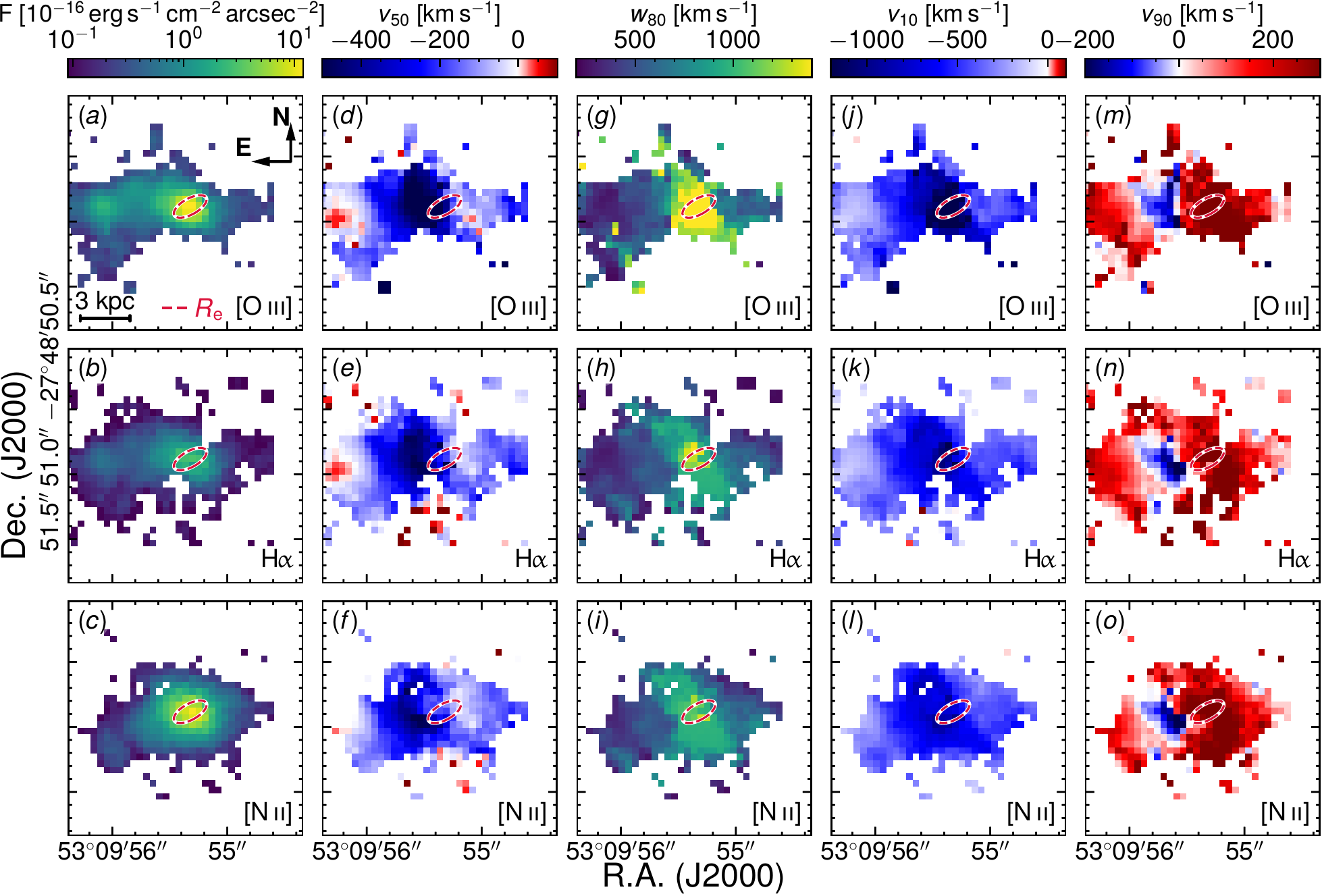}
  {\phantomsubcaption\label{f.emmaps.a}
   \phantomsubcaption\label{f.emmaps.b}
   \phantomsubcaption\label{f.emmaps.c}
   \phantomsubcaption\label{f.emmaps.d}
   \phantomsubcaption\label{f.emmaps.e}
   \phantomsubcaption\label{f.emmaps.f}
   \phantomsubcaption\label{f.emmaps.g}
   \phantomsubcaption\label{f.emmaps.h}
   \phantomsubcaption\label{f.emmaps.i}
   \phantomsubcaption\label{f.emmaps.j}
   \phantomsubcaption\label{f.emmaps.l}
   \phantomsubcaption\label{f.emmaps.m}
   \phantomsubcaption\label{f.emmaps.n}
   \phantomsubcaption\label{f.emmaps.o}
   }
  \caption{
  Flux and kinematic maps of the \OIIIall, \Halpha and \NIIall nebular lines.
  The three rows show \OIIIall (top), \Halpha (middle) and \NIIall (bottom).
  \textbf{Panels~\subref{f.emmaps.a}--\subref{f.emmaps.c}.} Flux maps, obtained
  fitting one or two Gaussians to the emission lines (see Methods).
  \textbf{Panels~\subref{f.emmaps.d}--\subref{f.emmaps.f}.} Non-parametric median
  velocity.
  \textbf{Panels~\subref{f.emmaps.g}--\subref{f.emmaps.i}.} Inter-percentile line
  width.
  \textbf{Panels~\subref{f.emmaps.j}--\subref{f.emmaps.l}.} 10\textsuperscript{th}
  percentile velocity.
  \textbf{Panels~\subref{f.emmaps.m}--\subref{f.emmaps.o}.} 90\textsuperscript{th}
  percentile velocity.
  For \target (whose continuum emission is highlighted by the red dashed ellipse), most of the emission is
  in a broad, blue-shifted component. Together with information from the
  BPT diagram (Fig.~\ref{f.bpt}), this is evidence for an AGN-powered outflow.
  }\label{f.emmaps}
\end{figure}

Spatially resolved emission-line kinematics (Fig.~\ref{f.emmaps}) display line
broadening in excess of 500~\kms, and centroids that are blueshifted relative to the 
stellar absorption (the stellar distribution is
traced by the red dashed ellipse in Fig.~\ref{f.emmaps}).
Most of the nebular flux comes from the central regions, but is
offset from the centre of the stellar emission and is elongated almost at right angle compared
to the stellar isophote (Fig.~\ref{f.emlines.c}). We also see a clear kinematic transition between this bright central
region and the more diffuse emission at larger radii. The central region has velocity offset
of -400~\kms, while the extended emission is at velocities of $\pm100~\kms$. Furthermore,
the central regions
have inter-percentile line widths \w80 in excess of 1,000~\kms.
The extent and elongation of the diffuse nebular emission suggest a merger origin,
but most of the flux is likely associated with the blueshifted component which we interpret 
as an outflow.
The large velocity offset and velocity width of the nebular lines -- and the fact that 
the contours of highest flux are directed along the galaxy minor axis
(Fig.~\ref{f.emlines.c} and Fig.~\ref{f.emmaps.a}--\subref{f.emmaps.c}) -- are evidence
for an outflow normal to the disc
\citep{cresci+maiolino2018, harrison+2018}. The fast outflow velocity, the low SFR,
and the AGN are evidence for SMBH-feedback driven outflows \citep{forsterschreiber+2019, nelson+2019,
concas+2022}.
We estimate a mass outflow
rate $\Mdotout[\OIII] = 0.14\text{--}2.9$~\MSun~\peryr (see Methods), meaning the
ionised-gas outflow is much lower than the SFR (Fig.~\ref{f.sfh.c}).

\begin{figure}
  \centering
  \includegraphics[width=1\textwidth]{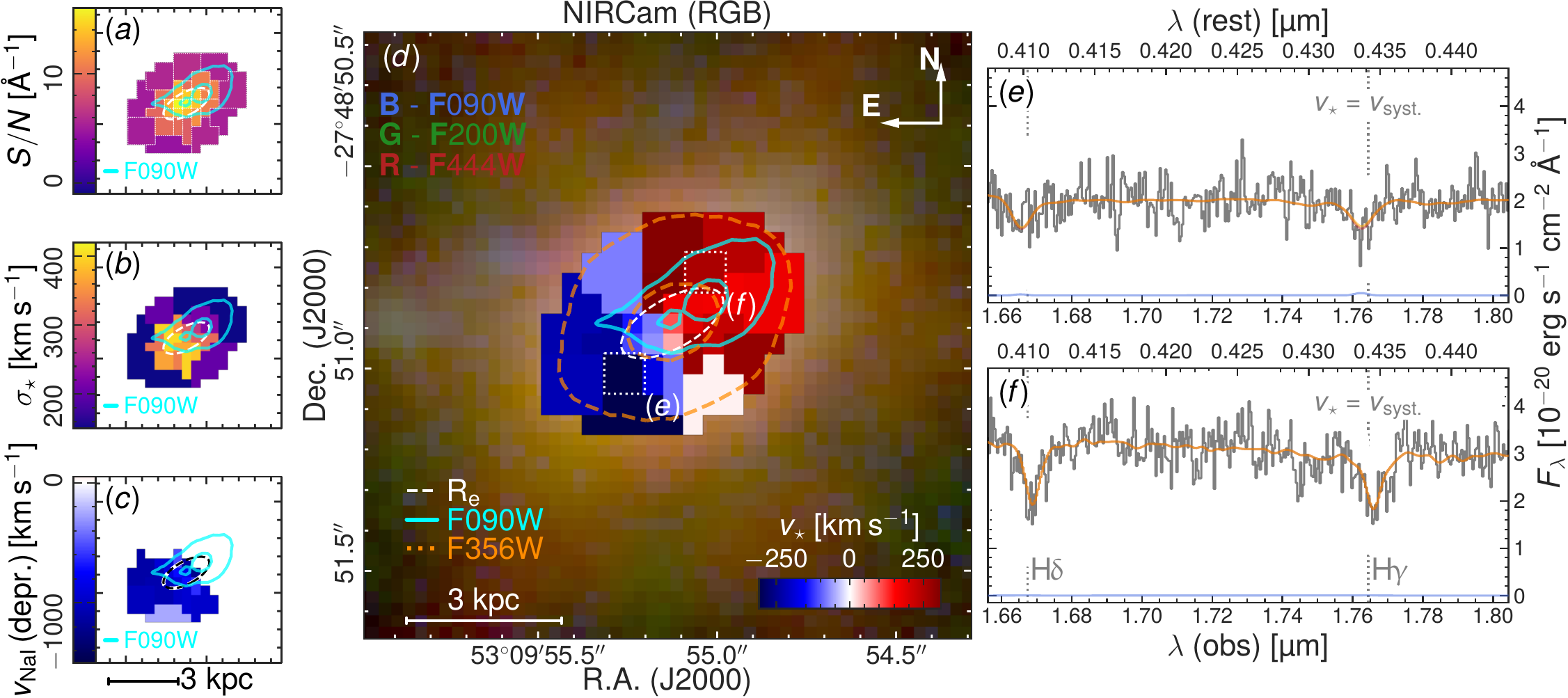}
  {\phantomsubcaption\label{f.maps.a}
   \phantomsubcaption\label{f.maps.b}
   \phantomsubcaption\label{f.maps.c}
   \phantomsubcaption\label{f.maps.d}
   \phantomsubcaption\label{f.maps.e}
   \phantomsubcaption\label{f.maps.f}
   }
  \caption{\textbf{Panels~\subref{f.maps.a}--\subref{f.maps.d}.} Voronoi-binned
  2-d maps of \target, showing the empirical S/N (panel~\subref{f.maps.a}),
  stellar velocity \vstar (panel~\subref{f.maps.d}),
  instrument-deconvolved velocity dispersion \sigmastar (panel~\subref{f.maps.b})
  and  neutral-gas velocity \vnai (panel~\subref{f.maps.c}). In these four
  panels the dashed ellipse traces the shape and \re of the best-fit S\'ersic
  model to the stellar continuum. Panel~\subref{f.maps.a} also shows the
  contours of the Voronoi bins (white dotted lines). Panel~\subref{f.maps.d}
  shows an enlarged region of the RGB image from Fig.~\ref{f.emlines.a} in the background, for
  comparison. Panel~\subref{f.maps.d} shows clear
  evidence for ordered rotation, with the rotation axis along the minor axis
  of the stellar distribution.
  \textbf{Panels~\subref{f.maps.e}--\subref{f.maps.f}.} Detail on the \Hdelta
  and \Hgamma stellar absorption features from two Voronoi bins (white
  dotted squares in panel~\subref{f.maps.d}), chosen to
  highlight the velocity offset that we attribute to stellar rotation. The
  line colours are the same as Fig.~\ref{f.emlines.e}--\subref{f.emlines.h}.
  }\label{f.maps}
\end{figure}

What sets \target apart from other high-redshift sources, however, is the clear
detection of spatially resolved, deep absorption features. These include the prominent hydrogen
absorption lines characteristic of 0.5-Gyr-old stellar populations (equivalent width
EW $\hda=6.6\text{--}8.1$~\AA and $\hga=4.1\text{--}6.8$~\AA; Figs.~\ref{f.emlines.g} and~\ref{f.maps.e}--\ref{f.maps.f}) and
\NaI resonant absorption (Fig.~\ref{f.emlines.h}). The Balmer absorption confirms the results
of the SED analysis (which is based on spectroscopy with much lower spectral resolution).
The \NaI absorption is offset both spatially and kinematically from the stars, therefore we
rule out a stellar origin. The highest \NaI EW translates into
a column density of $6\times10^{22}$~cm$^{-2}$, ten times lower than the
column density towards the X-ray AGN ($6\times10^{23}$~cm$^{-2}$, \citep{circosta+2019}).
The \NaI EW maps (Fig.~\ref{f.emlines.d}) show
that the absorption extends from the centre of the galaxy to the south-east, avoiding completely
the north-western half, where we detect the brighest rest-UV emission. This fact, combined with the 
blueshifted line-of-sight velocities ($\vnai=-(250\text{--}350)$~\kms), indicate that the
absorption is due to a neutral gas outflow, with a cone-like geometry possibly tilted with
respect to the stellar disc. We estimate the disc inclination from the best-fit
S\'ersic model of \citep{vanderwel+2014}, which gives a projected axis ratio
$q=0.44$ (we assume an intrinsic axis ratio $q_0=0.2$). After correcting for 
inclination by a factor $\cos\,i$, we obtain an outflow velocity of order $-1,000~\kms$ -- again too fast to be explained by star
formation alone. Assuming a maximal escape velocity of $\vesc = 1,200~\kms$, we infer a velocity
ratio $\vnai/\vesc = 0.8\pm0.2$, meaning the outflows are certainly capable of removing material
from the star-forming disc, with a fraction possibly escaping the galaxy altogether.

We assume an outflow extent $\Rout = 2.7~\kpc$, based on the largest distance between the
\NaI detection and the centre of the galaxy. With this \Rout and a thin-shell 
geometry, the mass outflow rate is $\Mdotout[\NaI] \approx
100~\MSun~\peryr$, with uncertainties of a factor of 3. This value is larger than the 
SFR -- particularly if we recall that the SFR is overestimated (Fig.~\ref{f.sfh.c}). 
This indicates that the neutral-gas outflow is capable of halting star formation by 
removing the necessary fuel.

In Fig.~\ref{f.maps} we show the stellar kinematics of \target. We measure a clear
velocity gradient between the east and west sides (cf. Fig.~\ref{f.maps.e}
and~\subref{f.maps.f}). Could this gradient be due to a major merger? In favour of this
hypothesis, the two sides of \target
have different continuum shapes, with the south-east side
(panel~\subref{f.maps.e}) displaying flatter continuum slope and
broader, less deep absorption. Rest-frame UV photometry shows multiple peaks,
and the stellar velocity-dispersion \sigmastar is highest in
the south east, rather than at the centre of the galaxy.
However, contrary to the merger expectation, both UV peaks are located on the same
(receding) region of the velocity map (cyan contours in Fig.~\ref{f.maps.d}),
therefore these two peaks are unlikely to trace two distinct galaxies merging.
In case of a merger, \sigmastar would be highest between the two galaxies, where
absorption lines of different redshift blend.
Moreover, the mass maps and the red-wavelength
NIRCam imaging show an elliptical morphology centred at -- or very near -- the
kinematic centre of the maps, which is also in contrast with the merger hypothesis
(Fig.~\ref{f.sfh.a}, and orange contours in Fig.~\ref{f.maps.d}, see also
Figs.~\ref{f.emlines.b}). 
Finally, the putative rotation axis is aligned with the minor
axis of the isophotes (and isodensity contours), as expected from regular
rotation.
Under the rotation hypothesis, the different spectral shapes
between the receding and approaching halves of the velocity map could be
due to a combination of more dust in the south east (where we also detect the
neutral-gas outflow, Figs.~\ref{f.emlines.d} and~\ref{f.maps.c}), a slightly younger
stellar population in the north west (Fig.~\ref{f.sfh.b}), and
stronger absorption-line infill due to nebular emission in the south east.
Most likely, all three
factors contribute to the observed spectral differences.

By applying the virial theorem to
the aperture velocity dispersion inside 1~\re, $\sigmaestar=356\pm36$~\kms, we find
a dynamical mass of \Mdynvalue (using the calibration of \citep{vanderwel+2022}), which
is comparable to \Mstar. In case of a major merger, both \sigmaestar and \re would 
be severely overestimated, meaning the true \Mdyn would be much lower than what inferred from the
virial theorem, in tension with \Mstar.
Based on all this evidence, we classify \target
as a regular (fast) rotator -- the most distant yet found from stellar kinematics.
An alternative possibility, is that \target is a system in an advanced stage of
merger, with the orbital angular momentum of the two progenitors locked in the
stars \citep{bois+2010}.
To measure the degree of rotation support, we use the \lambdar[\re] spin parameter, a 
proxy for the angular momentum per unit mass \citep{emsellem+2007}, defined by
\begin{equation}
    \lambdar[\re] \equiv \dfrac{
        \sum_{r<\re}^{} F_i r_i \vert v_i \vert}{
        \sum_{r<\re}^{} F_i r_i \displaystyle\sqrt{v_i^2 + \sigma_i^2}
        }
\end{equation}
where we calculate the radius $r_i$ of each spatial element along the best-fit 
ellipse \citep{brough+2017}, the aperture is the observed
isophote of semi-major axis one~\re, $F_i$ is the continuum flux, and $v_i$ and 
$\sigma_i$ are the stellar velocity and velocity dispersion.
We measure $\lambdar[\re][\text{(observed)}]=0.46$, then apply a model-based
point-spread function (PSF) correction (dependent on the galaxy size and 
light profile \citep{harborne+2020}) for a Gaussian PSF with full width at half 
maximum FWHM=0.09~arcsec, obtaining $\lambdar[\re] = 0.62\pm0.07$ (the uncertainties 
include the uncertainties due to the correction).

\begin{figure}
  \centering
  \includegraphics[width=1\textwidth]{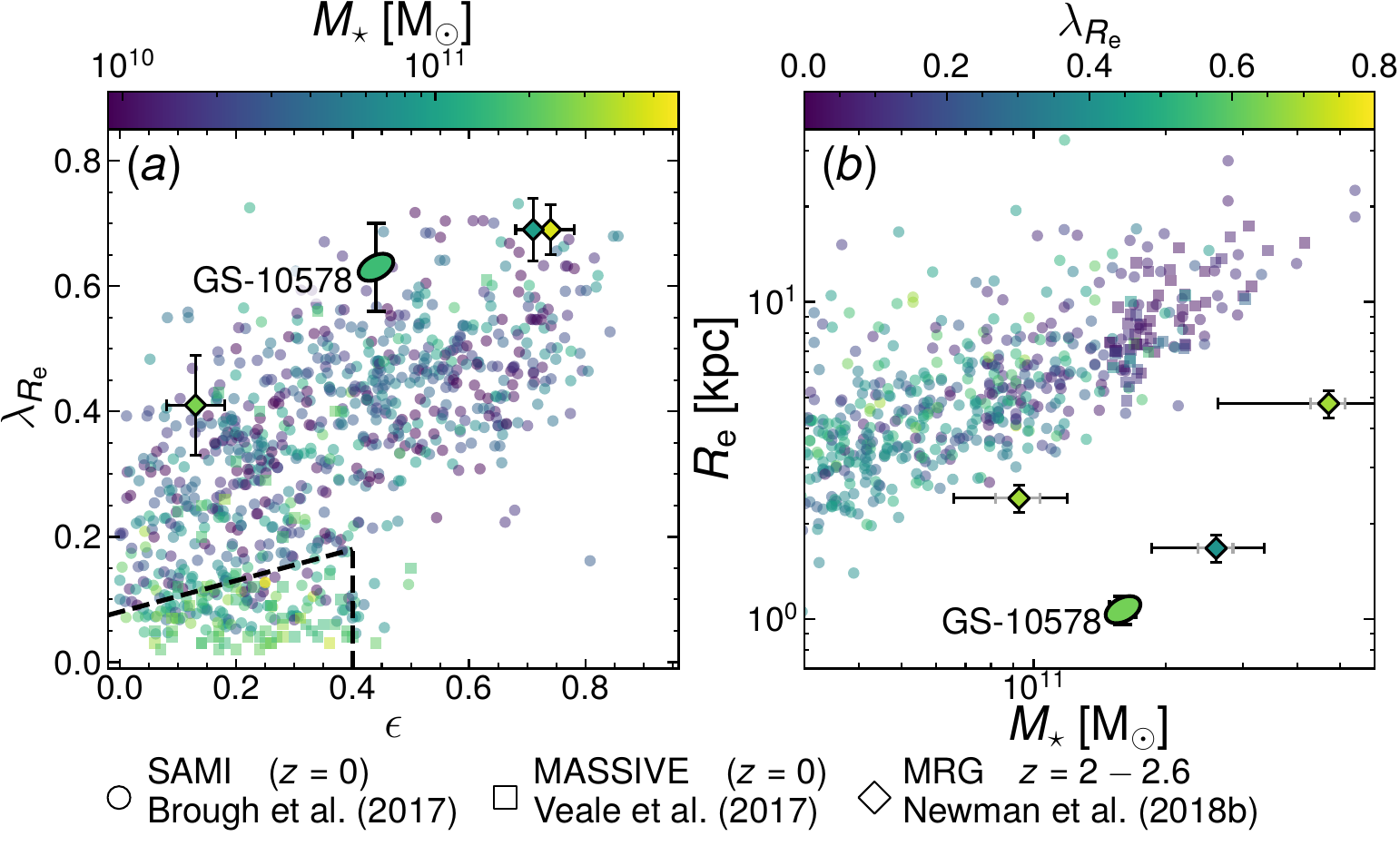}
  {\phantomsubcaption\label{f.leps.a}
   \phantomsubcaption\label{f.leps.b}
  }
  \caption{\textbf{Panel~\subref{f.leps.a}.} \lambdar[\re]--\eps diagram showing that
  -- interpreting the velocity field as stellar rotation -- \target is rotation
  supported, a so-called fast-rotator galaxy (slow rotators fall inside or near
  the dashed box, \citep{cappellari2016}). This is similar to other quiescent galaxies at high
  redshift (MRG sample, diamonds; \citep{newman+2018b}) and unlike local quiescent galaxies (SAMI sample, circles; \citep{brough+2017}), including the most massive 
  local quiescent galaxies (MASSIVE sample, squares; \citep{veale+2017}).
  \textbf{Panel~\subref{f.leps.b}.} Mass--size diagram, illustrating that \target
  is extremely compact; to join the $z=0$ distribution, our galaxy must grow in
  size by a factor of 10.
  }\label{f.leps}
\end{figure}

In Fig.~\ref{f.leps.a} we compare \target to other quiescent
galaxies. Based on its location on the \lambdar[\re]-\eps plane (where \eps is the
ellipticity of the isophote at 1~\re), \target is a clear
fast-rotator (FR) galaxy \citep{emsellem+2007}, i.e., it is consistent with being an
oblate spheroid with mild anisotropy, dominated by rotation support. FRs are rare 
among galaxies of similar mass in the local Universe \citep{trujillo+2009, 
comeron+2023}, but were much more common at high redshifts\citep{newman+2018b}.
Fig.~\ref{f.leps.b} shows \target on the mass-size plane: its compact size would again make
it a strong outlier at $z=0$ and even possibly at $z=2$. Taken together, 
Figs.~\ref{f.leps.a} and~\subref{f.leps.b} imply that galaxies like \target must
undergo significant dynamical and morphological evolution between $z=3$ and present.
This evolution has to decrease the angular momentum and increase the size -- all
while not exceeding the stellar mass of local massive galaxies. It has been
postulated that this kind of growth can only be achieved via gas-poor minor mergers
\citep{naab+2009, bezanson+2009}, because gas-rich merger trigger starbursts and disc
(re-)formation -- which would increase the light-weighted \lambdar[\re] parameter 
\citep{lagos+2018a, lagos+2022}. Our target is surrounded by three and up to five
faint, low-mass satellites, which may represent the fuel for the postulated size growth
\citep{suess_buddies_2023}.

\target represents a unique opportunity to study how the most massive galaxies in
the Universe became -- and stayed -- quiescent. Even though we cannot draw general
conclusions from a single target, we show that AGN feedback is capable of powering
neutral-gas outflows with high velocity and high mass loading, sufficient to interrupt
star formation by removing its cold-gas fuel. In contrast, the ionised-gas outflows
appear to have much lower mass loading and may be ineffective at halting star 
formation, in agreement with studies of other AGNs at high redshift.
It is unclear how much cold gas an AGN like the one active in \target may leave behind.
Non-detection from existing sub-mm observations of CO emission give a molecular-gas
mass fraction $f<10$~per cent \cite{circosta+2019}, lower than the value of
coeval star-forming galaxies \citep{tacconi+2020}, but not sufficiently constraining.
Current evidence from redshifts $z=1.5\text{--}2$ suggests that quiescent galaxies
have low (3~per cent) molecular-gas fractions \citep{whitaker+2021,williams_alma_2021}
-- an order of magnitude lower than coeval star-forming galaxies.
Any leftover gas may have low star-forming efficiency, perhaps due to
stabilisation against gravitational collapse and/or (AGN-induced) turbulence, similar 
to local PSB galaxies \citep{smercina_after_2022} and AGN hosts \citep{venturi+2021}.
The ISM/CGM around \target suggest that a combination of AGN feedback and mergers may
be providing the energy required to sustain this turbulence against dissipation.
The presence of a stellar disc suggests that feedback was `gentle' enough to preserve
the dynamical structure of \target throughout its quenching phase, therefore dynamical
transformation from disc-like FRs to spheroid-like SRs happens after quenching.
A census of the emission-line properties of high-redshift, massive
quiescent galaxies is essential to clarify if strong, ejective SMBH feedback is 
episodic or if it is a key mechanism for driving and maintaining quiescence.

\section{Methods}

\subsection{NIRSpec observations and data reduction}\label{m.data}

Our observations are part of the GA-NIFS survey, a large NIRSpec GTO 
survey targeting extended galaxies and AGNs at $z>3$ with the
NIRSpec Integral Field Spectroscopy (IFS, \citep{boker+2022}) mode
(PIs: S.~Arribas, R.~Maiolino).
Data for \target was obtained from programme PID~1216 (PI: N. L\"utzgendorf)
and consist of two disperser/filter combinations. Observations with the
prism/clear setup cover the full wavelength range of NIRSpec with spectral
resolution $R=30\text{--}300$ \citep{jakobsen+2022}, whereas the g235h/f170lp 
observations probe only 1.66--3.17~\mum but with higher resolution $R=2,700$.
Both observations had a \jwst position angle of -99\textdegree, resulting
in a NIRSpec position angle of 39.5\textdegree (measured north to east).
For prism/clear we used the {\sc IRS2RAPID} readout pattern, 33
groups per integration and 8 exposures, for a total of 1.1~h on source;
for g235h/f170lp we used the {\sc IRS2} readout pattern, 25 groups
per integration and 8 exposures, for a total of 4.1~h on source.
The exposures were dithered following the medium cycling pattern to
improve the PSF sampling and to marginalise against detector defects
and leakage/open shutters from the NIRSpec Micro-Shutter Assembly.

For the data reduction, we follow the procedure described in
\citep{perna+2023}; here we report a brief summary and highlight any
differences. We reduced the data 
with the publicly available {\sc jwst} pipeline, version 1.8.2, using 
static calibrations from the context file version 1014 (for g235h/f170lp)
and context file version 1068 (for prism/clear). The pipeline was patched
to include a number of corrections and improvements, as described in \citep{perna+2023}.
Compared to the default reduction, we used an additional correction for
the 1/f noise.
For g235h/f170lp, flux calibration was done in post-processing using data
from the standard star TYC~4433-1800-1 (PID 1128, observation 9).
To detect and flag outliers we used a custom algorithm, similar to
\citep{vandokkum+stanford2003}.
Their algorithm was developed specifically for (spatially) undersampled 
\hst images. However, NIRSpec/IFS data is undersampled only in the spatial
direction, therefore we calculated the 
derivative of the count-rate maps only along the detector x axis, which 
is the approximate direction of the dispersion axis (the trace curvature 
can be neglected on scales of two detector pixels).
The absolute value of the derivative was then normalised by the local flux
(or by 3$\times$ the noise, whichever was highest) and we rejected pixels
where the normalised derivative was higher than the
95\textsuperscript{th}-percentile of the resulting distribution.
For the cube-building stage, we used the drizzle algorithm to produce
cubes with a scale of 0.05 arcsec per spaxel, for both the prism/clear and
the g235h/f170lp datacubes.
For the PSF modelling (Method~III), we used a cube with 0.03~arcsec per spaxel.

We use publicly available photometry for additional flux calibration
(see \S~\ref{m.sed}).
At wavelengths 0.4--1.6~\mum we use \hst/ACS data from
GOODS \citep{giavalisco_goods_2004, illingworth+2016} and \hst/WFC3 IR 
data from CANDELS \citep{grogin+2011, whitaker+2019}; at wavelengths
0.9--4.8~\mum we use \jwst/NIRCam data from the \jwst Advanced Deep
Extragalactic Survey (JADES; PID~1180, PI: D.\ J.\ Eisenstein;
\citep{eisenstein_overview_2023, rieke_jades_2023}) and from the \jwst
Extragalactic Medium-band Survey JEMS (PID~1963, PIs: C.\ C.\ 
Williams, S.\ Tacchella and M.\ Maseda; \citep{williams_jems_2023}).

\subsection{Point-spread function determination}\label{m.psf}

We present three independent measurements of the shape and size of the
NIRSpec/IFS PSF. The first measurement uses observations of
a standard star with the g235h/f170lp grating/filter combination. The second
method uses a serendipitous star inside the NIRSpec/IFS field of view of \target. The
third method uses NIRCam imaging from the JADES and JEMS surveys
(for which the PSF is well understood) to infer the NIRSpec/IFS PSF.

\begin{figure}
  \centering
  \includegraphics[width=\textwidth]{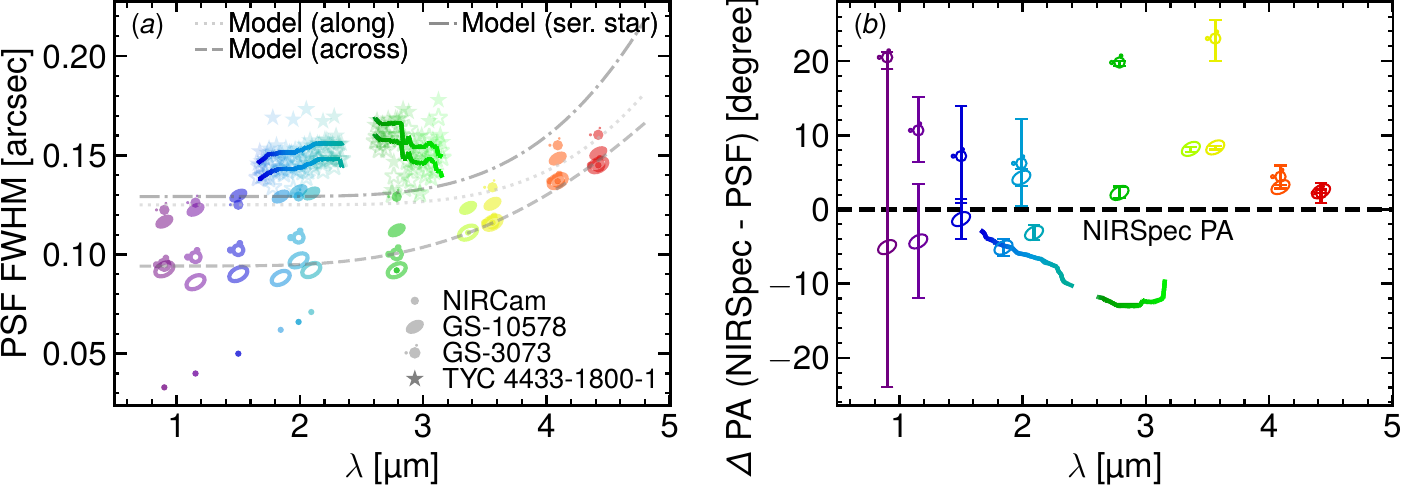}
  {\phantomsubcaption\label{f.psf.a}
   \phantomsubcaption\label{f.psf.b}}
  \caption{
  \textbf{Panel~\subref{f.psf.a}.} NIRSpec PSF
  as a function of wavelength, using three methods. Method~I (stars and solid
  lines) uses the calibration star TYC~4433-1800-1. Method~II (dot-dashed line)
  uses the serendipitous star inside the NIRSpec/IFS field of view for the observations
  of \target. Method~III uses NIRCam wide- and medium-band imaging to model the
  PSF with a 2-d Gaussian with orthogonal axes, with one axis closer to the slice direction (filled symbols) and one closer to across slices (empty symbols).
  The dots show the FWHM of the NIRCam
  empirical PSF for comparison.
  \textbf{Panel~\subref{f.psf.b}.} Position angle of the PSF major axis
  (points with error bars) compared to the NIRSpec position angle (dashed
  horizontal line). The major axis of the PSF is closer to the
  direction of the slices.
  }\label{f.psf}
\end{figure}

\textbf{Method~I -- Standard star observations.} We used the standard star
TYC~4433-1800-1 (V3=188\textdegree, PA=146.5\textdegree), reduced using the
same pipeline and context file as the galaxy
data, and modelled the PSF using the multi-Gaussian expansion algorithm
{\sc mgefit} \citep[MGE;][]{cappellari+emsellem2004}. We divide the datacube in
one hundred wavelength intervals between 1.66--3.17~\mum, then find the PSF
position angle from the second moment of the light distribution, and the
best-fit MGE model by minimising \textchi$^2$. The formal flux uncertainties on the 
wavelength-collapsed datacube are very small, and likely dominated by systematic
uncertainties. Our \textchi$^2$-minimisation uses therefore flux weighting instead
of error weighting.
The resulting FWHM is shown in Fig~\ref{f.psf.a}; filled and empty stars
are the model FWHM along the major and minor axis of the PSF, respectively; the
solid lines are the running median (note the detector gap at 
2.5~\mum). In panel~\subref{f.psf.b}
the solid line is the position angle of the PSF with respect to the NIRSpec
position angle; a value of 0\textdegree indicates that the PSF major axis has
the same direction as the slices long axes (hereafter: along slices; 90\textdegree
 is across slices).
We find the NIRSpec/IFS PSF to be preferentially aligned close to along slices.
At short wavelenghts, the PSF is larger than the NIRCam PSF (coloured dots),
and there seems to be a drop in FWHM around 3~\mum.

\textbf{Method~II -- Serendipitous Star.} For this method, we use the relatively
bright star inside the NIRSpec field of view (Fig.~\ref{f.emlines.a}; $F775W = 
25$~mag), modelled as a Gaussian. The prism/clear datacube for \target is summed
over six wavelength elements, and each synthetic image is fit independently. We then
model the resulting FWHM as a function of wavelength with the function:
\begin{equation}
    FWHM(\lambda) = ( 0.13 + 1.6 \cdot \lambda [\mum] \cdot \exp(-21.4 / \lambda [\mum])~\mathrm{arcsec}
\end{equation}
This model is shown as a dot-dashed grey line in Fig.~\ref{f.psf.a}; we remark that 
this determination may suffer from edge effects -- particularly at the longest
wavelengths.

\begin{figure}
  \centering
  \includegraphics[width=\textwidth]{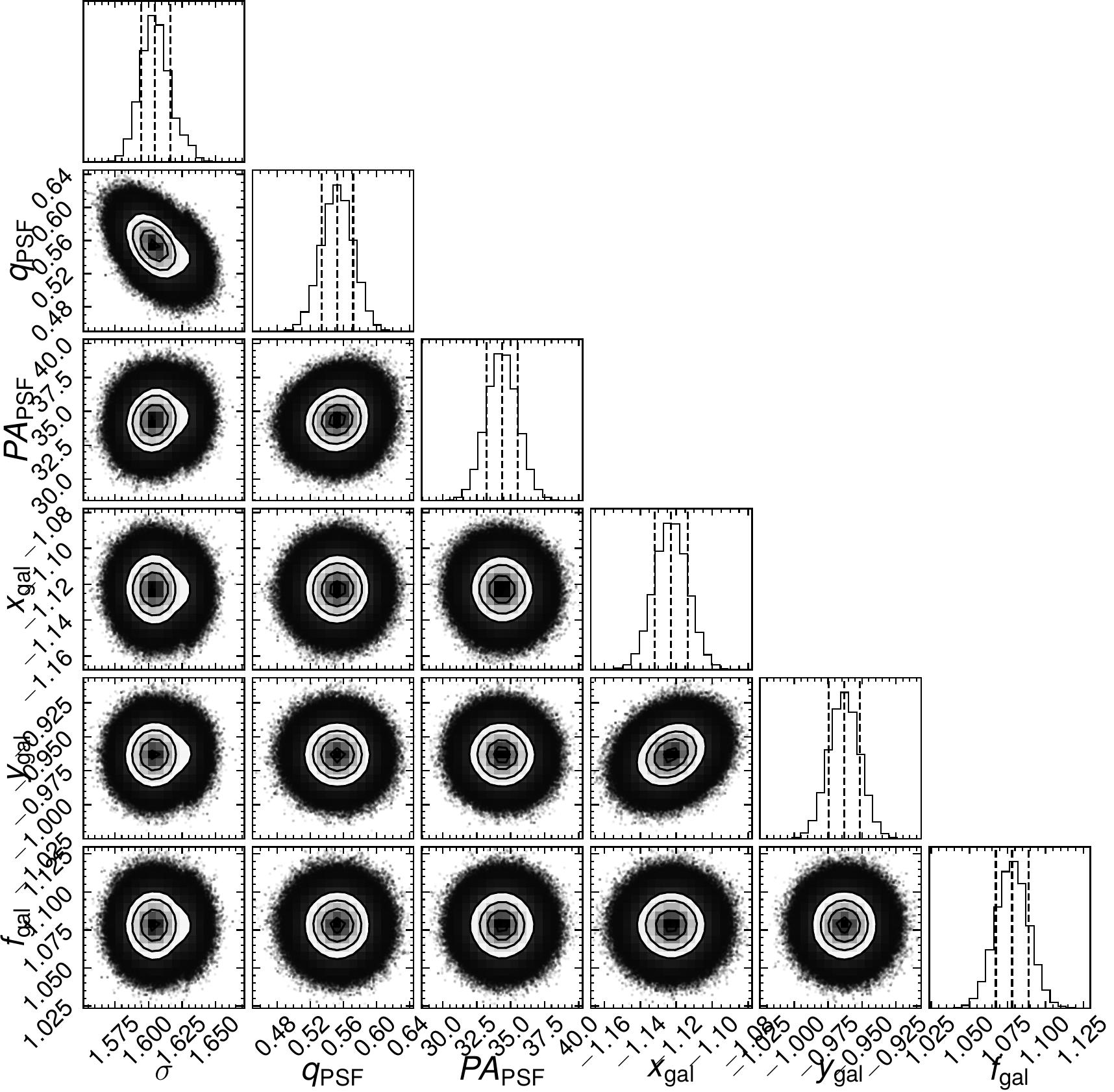}
  \caption{Example posterior distribution for our PSF estimation based on matching NIRCam imaging to synthetic NIRCSpec/IFS imaging (method~II). $\sigma$ is the semi-major axis standard deviation of the Gaussian kernel we apply to the NIRCAM
  F182M imaging to best match NIRSpec. The kernel has axis
  ratio $q=0.55$, but the NIRSpec/IFS PSF is the result of
  convolving this kernel with the NIRCam PSF. The kernel
  position angle is 34\textdegree, meaning the NIRSpec/IFS
  PSF is elongated preferentially along the IFS slices.
  }\label{f.psfchain}
\end{figure}

\textbf{Method~III -- Matching NIRCam imaging.} For this method, we select a
NIRCam wide- or medium-band filter, integrate the NIRSpec/IFS prism/clear
cube (0.03~arcsec/spaxel)
under the filter transmission curve to create a synthetic image in that band, then proceed to
find the kernel convolution that best matches the available NIRCam image to the
synthetic NIRSpec image. The NIRSpec/IFS PSF is then estimated as the convolution
of the NIRCam PSF with the best-fit kernel. The model consists of six free
parameters; the kernel is assumed to be a pixel-integrated, bivariate Gaussian, with 
three free
parameters $\sigma$, axis ratio $q$ and position angle.
In addition, we allow a small pixel adjustment to align the NIRCam and NIRSpec
images and a flux scaling factor. The model is optimised using the Markov-Chain
Monte-Carlo algorithm {\sc emcee} \citep{foreman-mackey+2013}. We use flat priors
for the kernel parameters ($0.003<\sigma<0.5$~arcsec, $-90<$position angle
$<90$\textdegree and $0<q<1$) and informative, Gaussian priors for matching the
NIRCam image position and flux (the pixel shifts are Gaussians with mean 0 and
standard deviation three pixels, and the flux scaling factor is a Gaussian with
mean equal to the mean flux ratio and standard deviation of 10 per~cent).
An example posterior distribution is shown in Fig.~\ref{f.psfchain}, which shows that
-- as for Method~I above -- the PSF is not circular.
Fig.~\ref{f.psf.a} shows the results from method~III as coloured ellipses, 
with the filled and empty symbols denoting the PSF FWHM along the major and minor
axes of the PSF, respectively. The position angle is once again
close to 0, i.e., the PSF tends to be aligned with the IFS slices 
(panel~\subref{f.psf.b}).
This method also finds a drop in the PSF FWHM between the NIRCam F210M and
F277W filters; the origin of this drop is unclear and we do not model it in
the subsequent analysis.
We tested that our results are consistent when modelling the convolution kernel
as a superposition of two Gaussians with the same axis ratio and position angle.
In addition to \target, we also model GS-3073, a AGN-host galaxy at $z=5.55$
\citep{uebler+2023}. The results are shown by the GS-3073-shaped markers, with filled
and empty symbols having the same meaning as for \target.
Using both \target and GS-3073, we model the PSF FWHM along the major axis (i.e.,
along slices) as
\begin{equation}
    FWHM(\lambda) = ( 0.12 + 1.9 \cdot \lambda [\mum] \cdot \exp(-24.4 / \lambda [\mum])~\mathrm{arcsec}
\end{equation}
(dotted grey line in panel~\subref{f.psf.a}) and along the minor axis (across slices) 
as
\begin{equation}
    FWHM(\lambda) = ( 0.09 + 0.2 \cdot \lambda [\mum] \cdot \exp(-12.5 / \lambda [\mum])~\mathrm{arcsec}
\end{equation}
(dashed grey line in panel~\subref{f.psf.a}).

Neither of the three methods adopted is ideal. For method~I, we are using observations
with a different number of dither positions than \target, which may affect the
image and PSF reconstruction; the limited number of dithers means we can only
reconstruct the image with a scale of 0.05~arcsec/spaxel, which may limit the
accuracy of the PSF modelling. For method~II, the serendipitous star is close to
the edge of the NIRSpec/IFS field of view, and may therefore suffer from
edge effects. For method~III, the starting NIRCam image has a different
PSF than the NIRSpec/IFS image, therefore, there is no guarantee that a Gaussian
(or two Gaussians) kernel can model the resulting image adequately. However, the fact
that the three methods agree on the elongated shape of the PSF leads credence to our
conclusion. Furthermore, the elongation of the PSF along the instrument slicers has
already been reported by other authors (Parlanti et al., in~prep., Perna et al.,
in~prep.), was already expected from the instrument design and had already been noted 
during ground-based evaluations (B\"oker,~T., priv.~comm.). 

When measuring the spin parameter \lambdar[\re], we apply an empirical correction
to obtain the intrinsic value from the observed measurement \citep{harborne+2020}.
We assume a PSF FWHM of 0.09~arcsec (at 2~\mum); a larger FWHM value would increase the
intrinsic \lambdar and make our results more significant.

\subsection{Redshift determination.}\label{m.redshift}

One would think that spectroscopy with spectral resolution of
$R=2,700$ ought to yield redshift uncertainties better than 10~\kms, i.e.,
a fraction of a pixel for well-sampled spectra (as is the case for NIRSpec).
However, our data consists of two kind of spectral features, both of which
are scarcely conducive of precise redshift determination. Nebular emission
consists of broad lines and of multiple narrow lines. Even when a precise
redshift can be determined, it is unclear which narrow lines can be used to
trace the redshift of the galaxy. For the stellar features, the physical
association with the systemic redshift is unambiguous, but all the absorption
features belong to the Balmer series of hydrogen and are therefore subject to
nebular-line infill. Line infill may consist of multiple components too.

To take into account the different ways the absorption lines can be filled we
use a Monte Carlo approach. We consider the central spectrum as the unweighted
sum of the spectra within the elliptical aperture of semi-major axis equal to
\re/2, axis ratio $q=0.75$ and position angle 111.5\textdegree; spaxels
partially inside the aperture are weighted according to the fractional overlap
between the spaxel and the ellipse. We then create one thousand random-noise
realisations of this spectrum by adding white noise from the uncertainty spectrum.

This procedure yields an ensemble of spectra that have 40~per cent lower S/N
than the actual data, but crucially preserve any systematics compared to e.g.,
bootstrapping based on a best-fit model spectrum. We fit each random
realisation with the same procedure used in \S~\ref{m.ppxf}, but we also use a
random starting guess for the redshift, chosen between $3.04<z<3.09$. The
resulting redshift distribution is bimodal, with the strongest peak at $z=3.06404$
and a secondary peak at $z=3.06568$ (the two distributions contain 63 and 37
per~cent of the total probability). We therefore adopt a systemic redshift
$z_\mathrm{sys} = 3.06404$. The random uncertainties (from a Gaussian fit to
the posterior distribution) are of order 0.0004, but this value is overestimated,
because we added random noise to the already noisy spectrum. However, the systematic 
uncertainty (defined as the distance between the strongest and secondary peak) is
much larger at 0.0016. The random and systematic
redshift uncertainties correspond to velocity uncertainties of 31~\kms and
120~\kms, respectively. Our final redshift determination is
$z_\mathrm{sys} = 3.06404\pm0.0004(\text{rand.})\pm0.0016(\text{syst.})$.

Spectroscopy in band 1 (g140m and g140h gratings) may remove the systematic 
uncertainty entirely, by
targeting the higher-order Balmer lines where the
stellar absorption is much stronger than the nebular
emission. However, even this region may present complex features (e.g., ISM
absorption with high equivalent width in the \CaII doublet, which overlap with
other stellar features (stellar \CaII and \Hepsilon).

\subsection{Spectral energy distribution modelling and star-formation history}\label{m.sed}

To study the physical properties of the stellar populations present in
\target, the prism/clear NIRSpec/IFS observations were analysed in a spaxel-by-spaxel basis (0.05~arcsec on a side) following a similar method to that described in \citep{2023ApJ...946L..16P}. Before comparing with models, the NIRSpec prism/clear data was PSF-matched using a Gaussian kernel to match the FWHM of the PSF at 4.5~\mum, 0.16~arcsec FWHM (see Fig.~\ref{f.psf.a}, dotted line)
The spectra were also re-calibrated in flux (with correction factors ranging from 0.86 at 1~\mum to 1.29 at 2.0--2.5~\mum) to match the integrated magnitude (in a 1.2~arcsec radius circular aperture) NIRCam wide- and medium-band data obtained by JADES \cite{eisenstein_overview_2023,rieke_jades_2023}. We considered a 5~per cent constant error added in quadrature to the formal observational values to account for the absolute flux calibration uncertainties. 

The spectrum for each spaxel counting with more than 10 wavelengths with $S/N>5$ was compared to a grid of stellar population models constructed with the \citep{bruzual+charlot2003} library, assuming a star formation history described by a delayed exponential characterised by a timescale $\tau$ (ranging from 1~Myr to 1~Gyr in 0.1~dex steps) and age $t_0$ (ranging from 1~Myr to the age of the Universe at the redshift of the galaxy, 2~Gyr). The stellar metallicity $Z$ was left as a free parameter, allowing to take all the discrete values provided by the   \citep{bruzual+charlot2003} library from 2~per cent solar to  2.5 times solar. Nebular (continuum and line) emission was taken into account as described in \citep{2003MNRAS.338..508P,2008ApJ...675..234P}. 
The attenuation of the stellar and nebular emission was modelled with a \citep{calzetti+2000}
law, with $A_V$ values ranging from 0 to 5 magnitudes in 0.1~mag steps. The (surviving, i.e., without including remnants and gas re-injected by winds and supernova explosions into the interstellar medium) stellar mass is obtained by scaling the (mass-normalised) stellar model to the spectrum (i.e., the stellar mass is not a directly fitted quantity).

The stellar population synthesis method includes a Monte Carlo algorithm to estimate uncertainties and degeneracies of the modelling (see \citep{2008ApJ...675..234P,2016MNRAS.457.3743D}), consisting in varying the data points according to their uncertainties, refitting the spectrum, and analysing the clusters of different solutions identified in the multi-dimensional parameter space defined by $\tau$, $t_0$, $Z$, and $A_V$. 

Based on this method, we constructed the stellar mass and mass-weighted age maps presented in Fig.~\ref{f.sfh.a}--\subref{f.sfh.b}. The mass-weighted age was calculated by integrating the star formation history multiplied by the time from 0 to the age $t_0$ of the stellar population. 

The integrated SFH and stellar mass evolution as well as the global spectral energy distribution, all presented in Fig.~\ref{f.sfh}, were constructed by adding all the spaxels included in the stellar population synthesis analysis. To account for low surface brightness pixels that were not included in the modelling, we applied an aperture correction obtained by comparing the flux of the analysed spaxels compared to the integrated photometric aperture.

\subsubsection{Modelling the panchromatic SED with {\sc cigale}}\label{m.cigale}

The panchromatic SED of the target was built by considering the multi-wavelength UV-to-NIR photometry available in \citep{Merlin2021} as well as the NIRCam photometry released by the JADES and JEMS collaborations \citep{eisenstein_overview_2023, rieke_jades_2023, williams_jems_2023}. The MIR-to-FIR photometry was taken from \citep{Shirley2021}. The counterpart in the different catalogs was found by using a positional matching radius of 1~arcsec. The SED was then analyzed by using the galaxy SED-fitting tool {\sc cigale} (Code Investigating GALaxy Emission; \citep{Boquien2019}). This code performs a multi-component fitting to separate the AGN contribution from the host galaxy emission. We used the following models to reproduce the different emission components: i) the stellar emission was modelled by adopting a delayed-$\tau$ (exponentially declining) SFH, the stellar population models of \citep{bruzual+charlot2003}, a Chabrier initial mass function (IMF; \citep{chabrier2003}), and solar metallicity. Dust attenuation was taken into account through the modified version of the \citep{Calzetti2000} curve available within {\sc cigale}. ii) The emission from dust heated by star formation was reproduced using the model library by \citep{Dale2014}, where we assumed an AGN contribution equal to 0 in order to treat it separately. iii) The AGN emission component was reproduced by using the models presented by \citep{Fritz2006}. We run {\sc cigale} by adopting the same input grid used by \citep{Circosta2018} and the {\sc cigale} version v2018.0.2.

\subsection{Stellar kinematics}\label{m.ppxf}

We present the first spatially resolved stellar kinematics beyond 
redshift $z=2.5$, and the first 2-d stellar kinematics beyond $z=0.8$. For these 
measurements we use \ppxf, which models
simultaneously the stellar continuum and nebular lines
with a non-linear \textchi$^2$ minimisation \citep{cappellari+emsellem2004,
cappellari2017, cappellari2022}, all while taking into account the instrument
line-spread function. The
stellar continuum is modelled as a (non-negative) linear superposition 
of simple stellar population (SSP) spectra. As input, we use a subset of
the SSPs library based on the MIST isochrones \citep{choi+2016} and C3K model
atmospheres \citep{conroy+2019}. The SSP templates span a 
logarithmically spaced grid of ages and metallicities, covering the 
ranges 10--2,500~Myr with 0.2~dex sampling and \ZH=-2.00--0.25~dex with 
0.25~dex sampling. The maximum age is set to 500~Myr older than the age 
of the Universe at the redshift of \target, to avoid edge effects (changing
this setting does not alter our conclusions).
The nebular emission and absorption lines are modelled as multiple Gaussians.
It is common
practice to also use additive Legendre polynomials when fitting stellar
kinematics \citep{vandesande+2017a}, but we found that \ppxf tended to fit a
strongly negative and almost featureless polynomial, akin to an
over-subtracted background as bright as the target, which  -- given our flux
calibration -- is unphysical.
We therefore used a 15\textsuperscript{th}-order multiplicative Legendre
polynomial to model residual flux-calibration issues.
For the line-spread function, we use a uniform spectral resolution of $R=2,700$,
using the \textit{nominal} resolution curve for g235h \citep{jakobsen+2022}.

We find that our results depend critically on the wavelength range adopted.
This is due to the already mentioned degeneracy between Balmer emission and
absorption (\S~\ref{m.redshift}).
For example, restricting the fit to the range where only \Hdelta and \Hgamma
are present, we measure a \sigmastar that is 25~per cent lower (in this case
the degree of the multiplicative polynomial is only 3).
It is common practice to assess this uncertainty by repeating the fit and
masking the Balmer lines \citep{newman+2018b}, but in our case these lines are
the only prominent stellar absorption features in the observed spectroscopic 
configuration. Therefore, as for the redshift, a conclusive test may require 
new observations at shorter wavelengths.

\subsubsection{Aperture-integrated kinematics}\label{m.ppxf.aperkin}

We use two sets of aperture-integrated kinematics: inside an ellipse of semi-
major axis equal to one~\re (for calculating the second moment of the
line-of-sight velocity distribution of the stars, \sigmaestar), and inside 
elliptical annuli (for use in the BPT diagram, Fig.~\ref{f.bpt}). In both 
cases, we use emission lines coupled in four sets called `kinematic 
components', which share the same line-of-sight velocity and intrinsic velocity 
dispersion (i.e., before convolution with the NIRSpec g235h line spread 
function). The first kinematic set consists of the Balmer
series of hydrogen \Hdelta--\Halpha (using the intrinsic line ratios appropriate
for Case~B recombination, electron density $n_e = 100~\mathrm{cm}^{-3}$ and
temperature $T_\mathrm{e}=10,000$~K; note the relative
line fluxes are then modified with a dust attenuation model),
\NIL, and the \OIall doublet. The second component consists of the
\OIIIall, \NIIall and \SIIall doublets. The third component consists of
\OIII and \NII. The fourth and last component is the resonant \NaI
$D_1$ and $D_2$ absorption features.
The \OIII, \OI and \NII doublets have line ratios fixed to the appropriate values
from atomic physics; the \SII doublet has line ratio constrained between 0.44 and 1.47
\citep{kewley+2019}, and the $D_2/D_1$ ratio of \NaI is constrained
between 0.89 and 1.94 \citep{poznanski+2012}.
For nebular emission only, we use the \citep{calzetti+2000} attenuation law.

From the elliptical aperture at 1~\re we find $\sigmaestar = 356\pm10$~\kms. The
uncertainties are likely dominated by systematics, so we consider the fiducial 
uncertainties to be 10~per cent.
Using the virial theorem calibration of \citep{vanderwel+2022} and the light
profile measurements from CANDELS \citep{grogin+2011} F160W imaging 
\citep{vanderwel+2014}, we
find a dynamical mass $\Mdyn = k(n)k(q) \sigmaestar^2 \re / \mathrm{G} =
(2.0\pm0.1)\times10^{11}$~\MSun, dominated by the scatter in the calibration,
which are of order 25~per cent. For a single-burst stellar population of age
1~Gyr, the surviving stellar mass fraction is 0.7 (assuming a
Chabrier initial mass function and solar metallicity).
Down-scaling the stellar mass formed by this factor, we obtain a
dynamical-to-stellar mass ratio of 70~per cent.

We also use the 1~\re apertures to measure the Balmer EW indices \hda and \hga,
using the Lick definition \citep{worthey1994}. For \hda, the Lick bands lie
partially outside the wavelength range of the g235h/f170lp datacube. We therefore
measure the \Hdelta absorption strength using the best-fit model. The resulting
values are $\hda=6.63\pm0.01$ and $8.06\pm0.08$~\AA (rest frame), where for the
second value we subtracted the nebular emission. The uncertainties are from
bootstrapping the uncertainties on the data. Clearly, the true stellar EW
must be higher than the non-subtracted measurement, which lies itself well into the
range that defines PSB galaxies ($\hda>4\text{--}5$~\AA, e.g., \citep{wu+2020}).
For \hga, we can measure the index from the data, and find respectively
$\hga=4.07\pm0.01$ and $6.83\pm0.09$~\AA for the unsubtracted and subtracted spectra.
Both values are again in the PSB regime.

\subsubsection{Spatially resolved stellar kinematics}\label{m.ppxf.spatkin}

Measuring spatially resolved stellar kinematics with \ppxf typically 
requires $S/N \gtrsim 10\text{--}15~\AA^{-1}$, unless specific
constraints are applied \citep{vandesande+2017a}. Our data has a continuum
$S/N$ as low as 3--5~$\AA^{-1}$, therefore we need to adopt a careful
approach. We measure spatially resolved stellar kinematics with two
different setups: Voronoi-binned data (Method~I) and spaxel-by-spaxel
measurements (Method~II). Briefly, Voronoi bins have the advantage of
higher $S/N$ \citep{cappellari+copin2003}, but for data with low spatial
resolution they risk averaging over regions with different stellar populations
and kinematics \citep{vandesande+2017a}; spaxel-by-spaxel fitting overcomes
the low-$S/N$ limitations by restricting the set of input templates
\citep{vandesande+2017a, foster+2021}.
The results in the article are based on Method~I. The
other method is described only briefly, to show that our findings do not
depend on the procedure adopted.

\textbf{Method~I -- Voronoi-binned data.} We measure the $S/N$ per
spaxel with a preliminary fit to each spaxel independently, using the
the best-fit template from the fit to the aperture spectrum as the only
input template (\ref{m.ppxf.aperkin}).
The $S/N$ is then measured from the standard deviation of the residuals
between the spaxel best-fit model and the data. Having determined the
$S/N$ empirically, we proceed to create the Voronoi bins using
{\sc vorbin} \citep{cappellari+copin2003}. We select a target $S/N = 
12~\AA^{-1}$. The resulting bins are outlined in Fig.~\ref{f.maps.a}
and are colour-coded by their empirical $S/N$, estimated from
the residuals of the \ppxf fit (see below); we note that in the
outer bins ($1 < R < 2~\re$) the empirical $S/N$ is lower than the
target $S/N$. This outcome is in fact expected from correlated noise,
which becomes more relevant as the average size of the Voronoi bins
increases outward from the galaxy centre.

To fit each spectrum, we use the same procedure as for the integrated
spectrum, but the choice of SSP spectra is restricted only to the
spectra that had non-zero weight in the linear combination forming
the best-fit aperture spectrum.

Moreover, the selection of Gaussian templates representing
nebular emission and absorption is also simplified compared to what we used
for the aperture spectrum, reflecting the simpler kinematics of the
spatially resolved spectra.
The first kinematic component consisting of the \Hdelta--\Halpha series,
the \OIIIall doublet, \NIL and the three doublets \OIall, \NIIall and
\SIIall. The second kinematic component consists of \OIIIall and \NIIall,
and the third kinematic component consists of \NaI absorption. All doublets
are fixed or constrained, as described for the aperture kinematics.
Example fits of Voronoi bins are shown in Fig.~\ref{f.maps.e} and~\subref{f.maps.f}.

\begin{figure}
    \centering
    \includegraphics[width=0.8\textwidth]{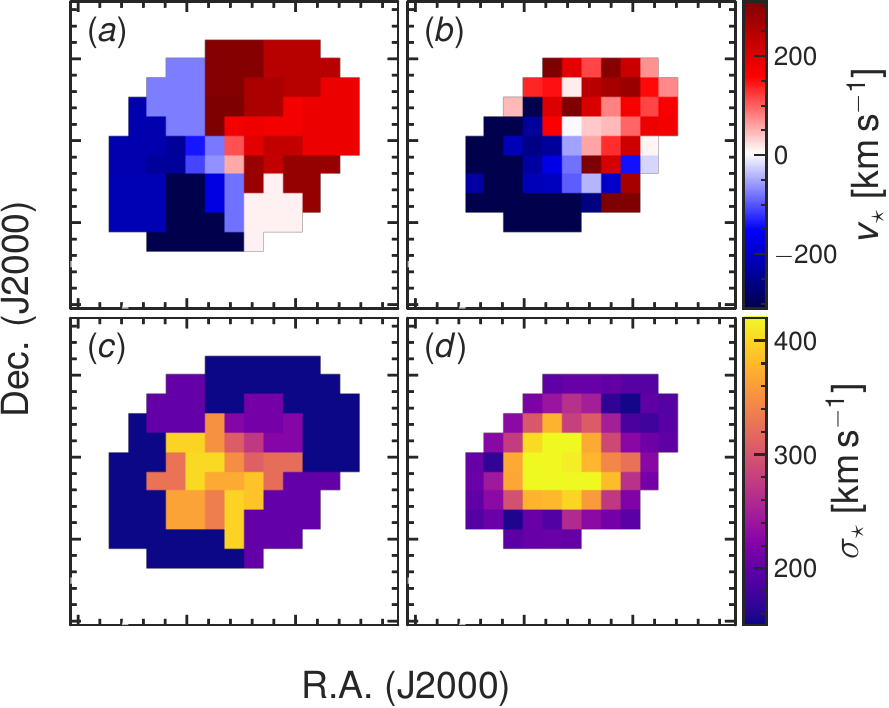}
    {\phantomsubcaption\label{f.spaxmap.a}
     \phantomsubcaption\label{f.spaxmap.b}
     \phantomsubcaption\label{f.spaxmap.c}
     \phantomsubcaption\label{f.spaxmap.d}}
    \caption{Comparison of the stellar kinematics between the fiducial method (based
    on Voronoi bins, Method~I; panels~\subref{f.spaxmap.a} and~\subref{f.spaxmap.c})
    and the alternative method (based on individual spaxel with constrained templates,
    Method~II; panels~\subref{f.spaxmap.b} and~\subref{f.spaxmap.d}). The two measurements
    show good agreement, but the spaxel-by-spaxel fits tend to find higher \sigmastar in
    the centre, possibly due to lower $S/N$ and smaller set of input templates for the fit.
    }
    \label{f.spaxmap}
\end{figure}

\textbf{Method~II -- Spaxel-by-spaxel data.} To measure spaxel-by-spaxel
kinematics, we use a threshold $S/N>3~\mathrm{pixel}^{-1}$. In general, such a low value is
considered sub-optimal, but if the templates are carefully pre-selected, it is
still possible to obtain accurate measurements \citep{vandesande+2017a}.
To optimise the templates selection, we first create 11
annular-binned spectra between 0 and 2.5~\re, spaced by 0.025 arcsec and with the same shape and position angle as
the observed stellar continuum. We then de-rotate the spectrum in each spaxel
according to the line-of-sight velocity measured in the relevant Voronoi bins;
this avoids the spectral broadening due to rotation.
The resulting bin spectra are fit with the same procedure as for the
Voronoi bins. From these fits we take the weights of 
the SSP templates and the
measured \sigmastar. We then proceed to fit the individual spaxels. Given a
spaxel, we identify the annular bins that the spaxel intersects and then also
consider two more bins (if they exist): the annulus adjacent to the innermost
selected annulus and the annulus adjacent to the outermost selected annulus.
The template selection is simply the union of all non-zero templates from the
best-fit annular spectra (typically 7--9 templates are selected). The velocity
dispersion is the average of the annular \sigmastar values, weighted by the
fractional overlap of the spaxel with the annuli.
We fit the spaxel spectrum using this restricted template selection and with
\sigmastar fixed to the average \sigmastar. If the resulting fit has $S/N<5$,
the fit is concluded and we proceed to another spaxel. If the resulting fit
has $S/N>5$, we repeat the fit leaving \sigmastar free between $\pm30$~per~cent
from the input \sigmastar.
The uncertainties are measured again using bootstrapping. The resulting maps are
shown in Fig.~\ref{f.spaxmap}, where we compare them with the Voronoi-binned maps.
The resulting velocity values are quantitatively different, but the results are
qualitatively the same. \sigmastar shows higher values in the central regions,
likely due to a combination of lower average $S/N$ and smaller template set
compared to the Voronoi bins. Using this method, we measure a spin parameter
$\lambdar[\re]=0.40$, which rises to 0.60 after the PSF correction; this latter
value is consistent with the measurement from Method~I.

\subsection{Emission-line measurements}\label{m.qubespec}

To measure the line-of-sight velocity and velocity width of the nebular emission 
lines, we used {\sc qubespec} algorithm. We modelled simultaneously \OIIIall, 
\Halpha and \NIIall. Each emission line was modelled with a single or a double 
Gaussian profile, with the redshift and FWHM of each Gaussian component being the 
same between the individual emission lines. We fixed the flux ratio of the \OIII 
and \NII doublets as described in \S~\ref{m.ppxf}.
We estimate the posterior probability distribution of both the single- and
double-Gaussian models with the Markov-Chain Monte-Carlo algorithm {\sc emcee}
\citep{foreman-mackey+2013}. We select the best model (either one or two
Gaussians per emission line) based on the Bayesian Information Criterion
\citep[BIC;][]{schwarz1978}. For each spaxel, we took a median spectrum of within 
the PSF ($\pm 2$ spaxel) to increase the $S/N$ of each fitted spectrum, while 
maintaining the diffraction-limited spatial resolution.

For each fitted spaxel, we calculated the \vperc{10}, \vperc{50}, \vperc{90} and \w80
as velocities of
10\textsuperscript{th}, 50\textsuperscript{th} and 90\textsuperscript{th} velocity
percentiles of the emission line and the velocity width of the line 80~per cent of the flux
(\vperc{90}-\vperc{10}). The systemic velocity of the galaxy (v$_{0}$) was set to the systemic
velocity of the stellar component (\S~\ref{m.redshift}). Despite each of the Gaussian
components fitted having the same redshift and FWHM, the final kinematics can be
different due to different flux ratios of the Gaussian components in each emission
line. The resulting flux and kinematic maps are presented in Fig.~\ref{f.emmaps}.

\subsection{Outflow measurements}\label{m.outflow}

To estimate the outflow rate of the ionised gas, we used the \OIII emission line. Because \Hbeta is
only seen in absorption, we calculated the mass of the outflowing ionised gas from the \OIII line as follows (see e.g. \citep{carniani+2015, kakkad+2021})
\begin{equation}
\Mout[\OIII] = 8\cdot 10^{7} \left(
\frac{1}{10^\mathrm{[O/H]-[O/H]_{\odot}}} \right) \left(
\frac{L_{\OIII}}{10^{44}~\mathrm{erg\,s^{-1}}} \right)\left(
\frac{\nelec}{500~\mathrm{cm}^{-3}}\right)^{-1}  M_{\odot} 
\label{eq:OIII_mass}
\end{equation}
where $L_{\OIII}$ is the line luminosity, \nelec is the mean electron density, and
where we assumed that the electron temperature is 10,000~K, all the oxygen is
ionised to O$^{2+}$, and the metallicity [O/H] of the outflowing material is
equal to the solar value [O/H]$_\odot$.
We further assume a uniformly filled biconical outflow, and use the equation from \citep{fiore+2017} as
\begin{equation}
\Mdotout = 3\cdot \vout \frac{\Mout}{\Rout}
\label{eq:mass_outflow}
\end{equation}
where \vout is the velocity of the outflows (defined as \vout=\vperc{10} = 1,200 km s$^{-1}$), \Mout
is the outflowing mass and \Rout is the radius of the outflow. The latter was
estimated in the \w80 map as the radius of the region where \w80$ > 600$~\kms 
\citep{harrison+2016,kakkad+2021}. This value corresponded to a radius
\Rout = 3.57~kpc. The largest source of uncertainty in this calculation is \nelec,
which 
can range from 500 to 10$^{4}$ cm$^{-3}$ (e.g., \citep{baron+netzer2019, davies+2020}). Therefore, we estimated the outflow rate for the range of density values, resulting in an outflow rate in ionised gas of 0.14--2.9~\MSun~\peryr. This
range is shown as the range of the cyan errorbars in Fig.~\ref{f.sfh.c}.\\

For the neutral gas outflows, we use the methods of \citep{cazzoli+2016}, based on the
\NaI absorption. These authors extend the methods of \citep{rupke+2005} to spatially
resolved spectroscopy. We assume a thin-shell, conical geometry orthogonal to the 
galaxy disc; \citep{maiolino+2012} remark that assuming a conical geometry with
uniform density
changes the outflow rate by a factor of 3. Our disc inclination is 66\textdegree,
from the axis ratio of the best-fit S\'ersic model \citep{vanderwel+2014}.
The outflow rate in the $k$\textsuperscript{th} spaxel is
\begin{equation}
    \Mdotout[\NaI][,k] = 11.5 \, \left(\dfrac{\Rout}{10~\kpc}\right)
    \left(C_f \dfrac{\Omega_k}{0.4}\right)
    \left(\dfrac{N_k}{10^{21}\,\mathrm{cm}^2}\right)
    \left(\dfrac{v_{\mathrm{out},k}}{200\,\kms}\right) \, \MSun \, \peryr
\end{equation}
where \Rout is the extent of the outflow, $\Omega_k$ is the subtended
solid angle, $C_f=0.37$ is the cloud covering fraction (estimated from
\citep{rupke+2005}),
$N_k$ is the gas column density, and $v_{\mathrm{out},k}$ is the outflow physical
velocity.
The extent of the outflow is $\Rout = 2.7~\kpc$,
estimated as the distance between the centre of the galaxy and the outermost
detected \NaI absorption, corrected for inclination. As noted by e.g.,
\citep{cazzoli+2016}, this size measurement relies on the presence of a sufficiently
bright background illumination to detect foreground \NaI absorption, therefore we
can consider \Rout a lower limit to the true extent of the outflow.
The solid angle subtended by each spaxel is calculated from the spherical shell
geometry.
$N_k$ is estimated from the equivalent width of \NaI using Eq.~1 from
\citep{cazzoli+2016}; our equivalent-width measurements are corrected for the
contribution due to the stellar atmospheric absorption.
The outflow velocity is taken from the {\sc ppxf} fits, and is the (absolute
value) of the centroid Gaussian-line velocity, corrected for the outflow
inclination by $\cos\,i$.

From this formula, summing over all spaxels where we detect blueshifted \NaI
absorption, we obtain $\Mdotout[\NaI] = 100$~\MSun~\peryr.\\

To estimate the escape velocity \vesc, we need the gravitational potential of
the galaxy, including its dark matter halo. Using the relation between stellar
mass and halo mass to infer the mass of the dark matter halo is impractical. This is
because -- for the stellar mass of \target\ -- the slope of the relation
is unfavourable for a precise inference.
We adopt the alternative method of assuming the gravitational potential
of the galaxy is an isothermal sphere, truncated at a radius $r_\mathrm{max}$
\citep{veilleux+2005, arribas+2014, uebler+2023}.
Using Eq~4 from \citep{uebler+2023} with $r_\mathrm{max}=100$~kpc,
we obtain an escape velocity $\vesc = 1,200\pm200$~\kms. The 
uncertainties were estimated with a simple Monte-Carlo analysis, where we have 
randomly drawn the various observables that go into the equation of \vesc
as follows. Aperture velocity dispersion \sigmaestar is drawn from the
normal distribution $\mathcal{N}(\mu, \sigma)$ with mean $\mu=356$ and
standard deviation $\sigma=36$~\kms (ten per~cent); \re is drawn from
$\mathcal{N}(0.14, 0.02)$ (units of kpc);
S\'ersic index $n$ is from the truncated normal distribution $\mathcal{T}(\mu, 
\sigma; \mu_0, \mu_1)$ with mean  $\mu=1.96$, $\sigma=0.4$ (twenty per~cent), 
and truncated between $\mu_0=1$ and $\mu_1=4$;
the $r_\mathrm{out}$ is from $\mathcal{T}(4.5, 1, 2.5, 6)$ (units of spaxels);
the galaxy intrinsic axis ratio $q_0$ is from $\mathcal{T}(0.2, 0.1, 0.1, 0.5)$,
and the projected axis ratio $q$ is from $\mathcal{T}(0.44, 0.1, 0.3, 0.5)$,
subject to $q\geq q_0$; $q$ and $q_0$ are used to infer the inclination $i$.

Most of the uncertainty on \vesc is due to the outflow radius $r_\mathrm{out}$
(60 per~cent), followed by \sigmaestar (30 per~cent) and \re (10 per~cent);
the other three variables account for a few per~cent each. Comparing our \vesc to
the map of de-projected \NaI velocities (Fig.~\ref{f.maps.c}), we find an average
ratio $\lvert \vnai \rvert / \vesc = 0.8\pm0.2$. Given the uncertainties on the
outflow geometry and escape velocity, we can conclude that at least part of the
neutral-gas outflows are able to escape the galaxy gravitational potential. At the
very least, the gas will be temporarily removed from the star-forming region.\\

Comparing the ionised and neutral outflow phases, we conclude that the
ionised-gas outflow rate is negligible, being up to two orders of magnitude lower than
the neutral-gas outflow rate. The mass loading factor of the neutral outflow is
orders higher than the recent SFR, therefore this
outflow is sufficient to \textit{stop} star formation. Similar events in the
recent past of \target may have caused the galaxy to become quiescent
(see SFH in \ref{f.sfh.b}).

\bigskip

\noindent{}
\textbf{Data availability}
The data that support the findings of this study are publicly available from the 
\href{Mikulski Archive for Space Telescopes}{https://archive.stsci.edu/}.

\noindent{}
\textbf{Code availability}
This work made extensive use of the freely available
\href{Debian GNU/Linux}{http://www.debian.org} operative system. We used the
\href{Python}{http://www.python.org} programming language
\citep{vanrossum1995}, maintained and distributed by the Python Software
Foundation. We further acknowledge direct use of
\href{astropy}{https://pypi.org/project/astropy/} \citep{astropyco+2013},
\href{dynesty}{https://pypi.org/project/dynesty/} \citep{speagle2020},
\href{fsps}{https://github.com/cconroy20/fsps} \citep{conroy+2009, conroy_gunn_2010},
\href{matplotlib}{https://pypi.org/project/matplotlib/} \citep{hunter2007},
\href{numpy}{https://pypi.org/project/numpy/} \citep{harris+2020},
\href{prospector}{https://github.com/bd-j/prospector} \citep{johnson+2021}
\href{python-fsps}{https://pypi.org/project/dynesty/} \citep{johnson_pyfsps_2023},
and \href{scipy}{https://pypi.org/project/scipy/} \citep{jones+2001}.

\backmatter

\section*{Acknowledgements}
{\small 
FDE, TJL, RM and JS and acknowledge support by the Science and Technology Facilities Council (STFC), by the ERC Advanced Grant 695671 ``QUENCH'', and by the UKRI
Frontier Research grant RISEandFALL; RM is further supported by a research professorship from the Royal Society.
PGP-G acknowledges support grants PGC2018-093499-B-I00 and PID2022-139567NB-I00 funded by Spanish Ministerio de Ciencia e Innovación MCIN/AEI/10.13039/501100011033, FEDER, UE.
SA, BRP, and MP acknowledge support from the research project PID2021-127718NB-I00 of the Spanish Ministry of Science and Innovation/State Agency of Research (MICIN/AEI). MP is further supported by the Programa Atracci\'on de Talento de la Comunidad de Madrid via grant 2018-T2/TIC-11715.
H{\"U} gratefully acknowledges support by the Isaac Newton Trust and by the Kavli Foundation through a Newton-Kavli Junior Fellowship.
AJB, JC and GCJ acknowledge funding from the ``FirstGalaxies'' Advanced Grant from the European Research Council (ERC) under the European Union's Horizon 2020 research and innovation programme (Grant agreement No. 789056).
SC, EP and GV acknowledge support by European Union's HE ERC Starting Grant No. 101040227 - WINGS.
GC acknowledges the support of the INAF Large Grant 2022 ``The metal circle: a new sharp view of the baryon
cycle up to Cosmic Dawn with the latest generation IFU facilities''
IL acknowledges support from PID2022-140483NB-C22 funded by AEI 10.13039/501100011033 and BDC 20221289 funded by MCIN by the Recovery, Transformation and Resilience Plan from the Spanish State, and by NextGenerationEU from the European Union through the Recovery and Resilience Facility.
BER acknowledges support from the NIRCam Science Team contract to the University of Arizona, NAS5-02015. The authors acknowledge use of the lux supercomputer at UC Santa Cruz, funded by NSF MRI grant AST 1828315.
}


\newcommand\aap{A\&A}                
\let\astap=\aap                          
\newcommand\aapr{A\&ARv}             
\newcommand\aaps{A\&AS}              
\newcommand\actaa{Acta Astron.}      
\newcommand\afz{Afz}                 
\newcommand\aj{AJ}                   
\newcommand\ao{Appl. Opt.}           
\let\applopt=\ao                         
\newcommand\aplett{Astrophys.~Lett.} 
\newcommand\apj{ApJ}                 
\newcommand\apjl{ApJ}                
\let\apjlett=\apjl                       
\newcommand\apjs{ApJS}               
\let\apjsupp=\apjs                       
\newcommand\apss{Ap\&SS}             
\newcommand\araa{ARA\&A}             
\newcommand\arep{Astron. Rep.}       
\newcommand\aspc{ASP Conf. Ser.}     
\newcommand\azh{Azh}                 
\newcommand\baas{BAAS}               
\newcommand\bac{Bull. Astron. Inst. Czechoslovakia} 
\newcommand\bain{Bull. Astron. Inst. Netherlands} 
\newcommand\caa{Chinese Astron. Astrophys.} 
\newcommand\cjaa{Chinese J.~Astron. Astrophys.} 
\newcommand\fcp{Fundamentals Cosmic Phys.}  
\newcommand\gca{Geochimica Cosmochimica Acta}   
\newcommand\grl{Geophys. Res. Lett.} 
\newcommand\iaucirc{IAU~Circ.}       
\newcommand\icarus{Icarus}           
\newcommand\japa{J.~Astrophys. Astron.} 
\newcommand\jcap{J.~Cosmology Astropart. Phys.} 
\newcommand\jcp{J.~Chem.~Phys.}      
\newcommand\jgr{J.~Geophys.~Res.}    
\newcommand\jqsrt{J.~Quant. Spectrosc. Radiative Transfer} 
\newcommand\jrasc{J.~R.~Astron. Soc. Canada} 
\newcommand\memras{Mem.~RAS}         
\newcommand\memsai{Mem. Soc. Astron. Italiana} 
\newcommand\mnassa{MNASSA}           
\newcommand\mnras{MNRAS}             
\newcommand\na{New~Astron.}          
\newcommand\nar{New~Astron.~Rev.}    
\newcommand\nat{Nature}              
\newcommand\nphysa{Nuclear Phys.~A}  
\newcommand\pra{Phys. Rev.~A}        
\newcommand\prb{Phys. Rev.~B}        
\newcommand\prc{Phys. Rev.~C}        
\newcommand\prd{Phys. Rev.~D}        
\newcommand\pre{Phys. Rev.~E}        
\newcommand\prl{Phys. Rev.~Lett.}    
\newcommand\pasa{Publ. Astron. Soc. Australia}  
\newcommand\pasp{PASP}               
\newcommand\pasj{PASJ}               
\newcommand\physrep{Phys.~Rep.}      
\newcommand\physscr{Phys.~Scr.}      
\newcommand\planss{Planet. Space~Sci.} 
\newcommand\procspie{Proc.~SPIE}     
\newcommand\rmxaa{Rev. Mex. Astron. Astrofis.} 
\newcommand\qjras{QJRAS}             
\newcommand\sci{Science}             
\newcommand\skytel{Sky \& Telesc.}   
\newcommand\solphys{Sol.~Phys.}      
\newcommand\sovast{Soviet~Ast.}      
\newcommand\ssr{Space Sci. Rev.}     
\newcommand\zap{Z.~Astrophys.}       
\bibliographystyle{ancillary/sn-mathphys}
\bibliography{ancillary/astrobib}

\end{document}